\documentclass{aa}  

\usepackage{graphicx}
\usepackage{natbib}     
\usepackage{hyperref} 
\usepackage{txfonts}
\newcommand{\slfrac}[2]{\left.#1\middle/#2\right.}

\begin{document}

   \title{A comparative analysis of the outer-belt primitive families}

   \author{M. N. De Pr\'a\inst{1},
          N. Pinilla-Alonso \inst{1},
          J. Carvano \inst{5},
          J. Licandro \inst{2,3},
          D. Morate \inst{5},
          V. Lorenzi \inst{4,2},
          J. de Le\'on \inst{2,3},
          H. Campins \inst{6},
          T. Moth\'e-Diniz \inst{7}}
          
   \authorrunning{M. N. De Pr\'a et al.} 
   
   \institute{Florida Space Institute, University of Central Florida, Florida, USA\\
              \email{mariodepra@ucf.edu},
              \and
             Instituto de Astrof\'isica de Canarias, C/V\'ia Láctea s/n, 38205 La Laguna, Spain
             \and
             Universidad de La Laguna (ULL), E-38205, La Laguna, Spain
             \and
             Fundaci\'on Galileo Galilei - INAF, Rambla José Ana Fern\'andez P\'erez, 7, 38712 Bre\~na Baja, Santa Cruz de Tenerife, Spain
             \and
             Departamento de Astrof\'isica, Observat\'orio Nacional, Rio de Janeiro, 20921-400, Brazil
             \and
             Department of Physics, University of Central Florida, Florida,  USA
             \and
             Faculty of Natural Sciences, Norwegian University of Science and Technology (NTNU), 7491, Trondheim, Norway
             }

   \date{April 08, 2020 }

% \abstract{}{}{}{}{} 
% 5 {} token are mandatory
 
  \abstract
   {Asteroid families are witnesses to the intense collisional evolution that occurred on the asteroid belt. The study of the physical properties of family members can provide important information about the state of differentiation of the parent body and provide insights into how these objects were formed. Several of these asteroid families identified across the main belt are dominated by low-albedo, primitive asteroids. These objects are important for the study of Solar System formation because they were subject to weaker thermophysical processing and provide information about the early conditions of our planetary system.}
   {We aim to study the diversity of physical properties among the Themis, Hygiea, Ursula, Veritas, and Lixiaohua families.}
   {We present new spectroscopic data, combined with a comprehensive analysis using a variety of data available in the literature, such as albedo and rotational properties.}
   {Our results show that Themis and Hygiea families, the largest families in the region, present similar levels of hydration. Ursula  and Lixiaohua families are redder in comparison to the others and present no sign of hydrated members based on the analysis of visible spectra. Conversely, Veritas presents the highest fraction of hydrated members.}
   {This work demonstrates a diverse scenario in terms of the physical properties of primitive outer-belt families, which could be associated with dynamical mixing of asteroid populations and the level of differentiation of the parental body.}


   \maketitle
%
%-------------------------------------------------------------------

\section{Introduction}

        Asteroid families are remnants from the main-belt collisional history, and are identified by clusters in the proper orbital parameters \citep{1918AJ.....31..185H, 1992CeMDA..54..207Z, 2015aste.book..297N}. Members of an asteroid family consist of fragments of a parent body that were disrupted after experiencing an energetic collision in the past. The study of physical properties among family members, such as composition, degree of aqueous alteration, and spectral diversity, can provide important information about the state of differentiation of the parent body  \citep{2002aste.book..633C}. Knowledge of the internal structure of asteroids is particularly difficult to obtain, yet is critical to understand how, when, and where these objects were formed and to shed light on the formation and evolutionary history of the  Solar System.
        
        Several asteroid families identified across the main belt are dominated by low-albedo, primitive asteroids \citep{2015aste.book..297N, 2015PDSS..234.....N}. These objects are of particular importance to the study of the  formation of the Solar System because they were subject to low thermophysical processing or none at all, and can provide information about the early conditions of our planetary system.  Furthermore, investigations of the surface composition of these objects show that they can contain hydrated minerals, water ice, and organic materials \citep{1994Icar..111..456V, 2014Icar..233..163F, 2010Natur.464.1320C, 2010Natur.464.1322R,  2011A&A...525A..34L, 2012Icar..219..641T, 2015aste.book...65R, 2019JGRE..124.1393R}. These discoveries corroborate the hypothesis that  primitive asteroids are a potential source of Earth's water and organic material \citep{2000M&PS...35.1309M}. 
        
        Advances in the characterization of primitive asteroids show that they present almost featureless visible and near-infrared (NIR) spectra, with a variety of surface colors from blue to red (for a review see \citealt{2015aste.book...13D}). A common feature observed in several of these objects across the asteroid belt is the 0.7 $\mu$m absorption band, characteristic of hydrated minerals \citep{1994Icar..111..456V, 2003Icar..161..356C, 2014Icar..233..163F}. The number of objects presenting this feature peaks at the central belt and decreases at the Cybele region, delimiting the hydration zone \citep{2012Icar..221..744R, 2014Icar..233..163F, 2018Icar..311...35D}. As part of the PRIMitive asteroid spectroscopic survey (PRIMASS), we studied the visible spectroscopy of eight primitive asteroids families in the inner belt and identified  a mix of anhydrous and hydrated families in that region, where the percentage of hydrated members in a family can reach nearly 80\% \citep{2016Icar..266...57D, 2016A&A...586A.129M, 2018A&A...610A..25M, 2019A&A...630A.141M}. 
        
\begin{figure}[h!]
                \centering
                \includegraphics[scale=0.67]{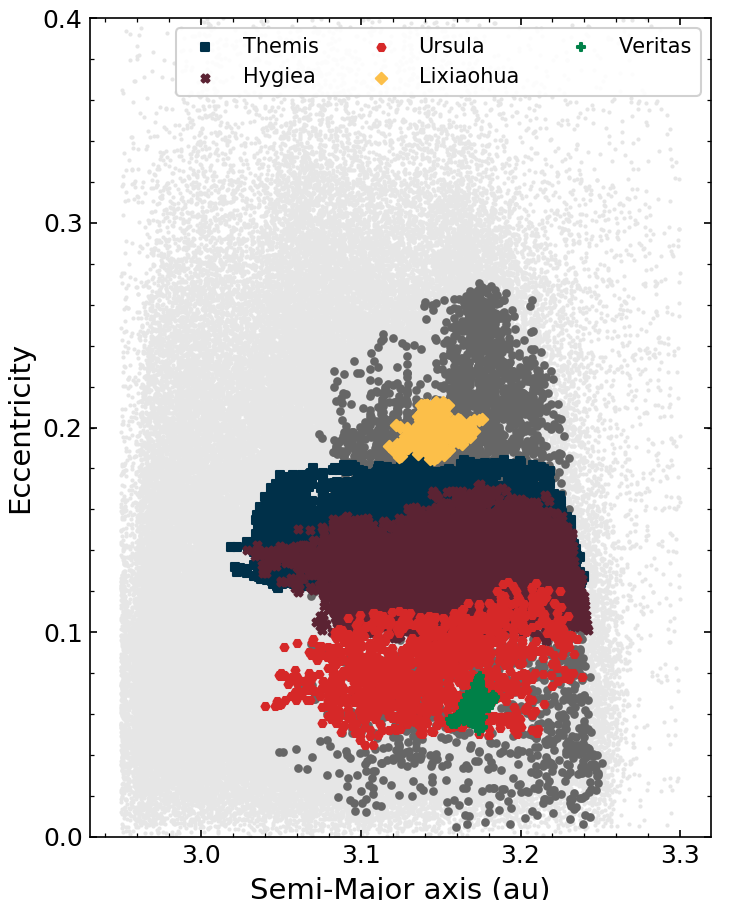}
                \caption{Orbital properties of large outer-belt asteroid families. The studied families are shown in colors, while those that are not studied in this work are represented by dark gray dots. The light gray dots represent the background.}
                \label{fig:orbital}
        \end{figure}

        Primitive asteroid families are observed across the whole main belt, and they become the dominant population in the outer belt where most of their mass is contained \citep{2014Natur.505..629D}. \cite{2015aste.book..297N} identified nine large ($N>300$ members) primitive families in this region (Fig. \ref{fig:orbital}): Themis, Hygiea, Ursula, Veritas, Lixiaohua, Alauda, Theobalda, Euphrosine, and Meliboea.
        
                \begin{table*}[h!]
                \centering
                \footnotesize
                \setlength\tabcolsep{3.0pt}
                \begin{tabular}{lcccccccccc}
                        \hline
                        Asteroid & a & e & i & Age & Members  & Spectra & SDSS & MOVIS & WISE & ALCDEF \\
                        Family & (au)&  & ($^\circ$)  & (Gyr)& (N$^\circ$) & (N$^\circ$) &  (N$^\circ$)& (N$^\circ$) & (N$^\circ$) & (N$^\circ$) \\
                        \hline
                        Themis    & 3.135 & 0.153 & 10.149  & $2.5\pm1.0$ & 4782 & 59 & 774 & 284  & 2312 & 443\\
                        Hygiea    & 3.142 & 0.136 & 1.084  & $2.0\pm1.0$  & 4854 & 25 & 783 & 182 & 2234 &149\\
                        Veritas   & 3.174 & 0.066 & 5.103  & $0.0083 \pm 0.0005$ & 1294  & 17 & 234 & 34 & 715 & 41\\
                        Ursula    & 3.129 & 0.092 & 9.164  & $<3.5$ & 1466 & 7 & 266 & 71  & 957 &70\\
                        Lixiaohua & 3.153 & 0.197 & 3.215  & $0.15 \pm 0.05$ & 756 & 35 & 110 & 24 & 514 & 28 \\
                        \hline
                \end{tabular}
                \caption{Outer belt families identified by \cite{2015aste.book..297N} and studied in this work. See text for details.}

                \label{tab1}
        \end{table*}
        
        The first detection of water ice in the asteroid belt was made on (24) Themis, the largest member of its homonymous family \citep{2010Natur.464.1320C, 2010Natur.464.1322R}. Recent works proposed that water ice is also present in large outer-belt asteroids \citep{2012Icar..219..641T, 2019JGRE..124.1393R}. Additionally, the first observed main-belt comet (MBC), an active asteroid with proposed sublimation-driven activity, is also a member of the Themis family. \citep{2004AJ....127.2997H, 2006Sci...312..561H}. \cite{2018AJ....155...96H} proposed that there are seven MBCs which are linked to four families in outer belt: Themis, Lixiahoua, Alauda, and Theobalda.
        
        The Themis family is one of the most highly studied families in the literature, in which both water ice and hydrated minerals have been observed among family members. \cite{2010GeoRL..3710202C} and \cite{2016A&A...586A..15M} interpreted these results as the product of a partial differentiation of the Themis parent body. The spectroscopic properties of the Lixiaohua family were studied in  \cite{de2020spectroscopic}, and although it is also a MBC-bearing family, it was found to present distinct properties from the Themis group, which can be interpreted as a difference in composition and/or differentiation level.
         
        The aforementioned studies of outer-belt objects suggest that the hydrated to icy transition is likely to occur at the outer belt. In this work, we performed a comparative study of the physical properties among the outer-belt families by means of new spectroscopic data combined with data gathered from public catalogs. In Section \ref{sec:obs} we describe the spectroscopic observations and data-reduction techniques. The analysis of these data is presented in Section \ref{sec:ana}, and we include the data available from public catalogs to extend the analysis to study the albedo, parent body size, and rotational properties of these families. The results for each family are presented in Section \ref{sec:res}. In Section \ref{sec:dis}, we compare results across the families and discuss the implications of these findings.

        \section{Observations and data reduction}
        \label{sec:obs}
        
        We collected low-resolution spectra for 29 asteroid members of the  Themis, Hygiea, and Veritas families (Table \ref{table:visobs}, Figs. \ref{fig:themis}, \ref{fig:hygiea}, \ref{fig:veritas} ). The data were obtained using the Goodman High-Throughput Spectrograph (GTHS) at the 4.1m SOAR telescope on Cerro-Pach\'on, Chile. We used a setup with grating of 300 lines/mm and the slits of 1.03'' in 2011, and 1.68'' in 2012 with no second-order blocking filter, which provided an effective spectral interval of 0.4-0.87 $\mu$m. Observations were made on a total of six nights across the semesters of 2011A and 2012B. We also obtained two sequences of calibration quartz lamps immediately before and after target acquisition in order to reduce systematic errors. At least one solar analog was observed at different air masses during each night. 
        
        We used quartz lamps for the flat-field correction of the images, while the HgAr lamps were used for the wavelength calibration. The reduction was made using standard techniques: images were bias- and flat-field corrected and the spectra were extracted, background subtracted, and wavelength calibrated in sequence. This procedure was repeated for the three sub-exposures, and then averaged to produce a final spectrum for each target.

        \begin{figure*}[t!]
                \centering
                \includegraphics[scale=0.55]{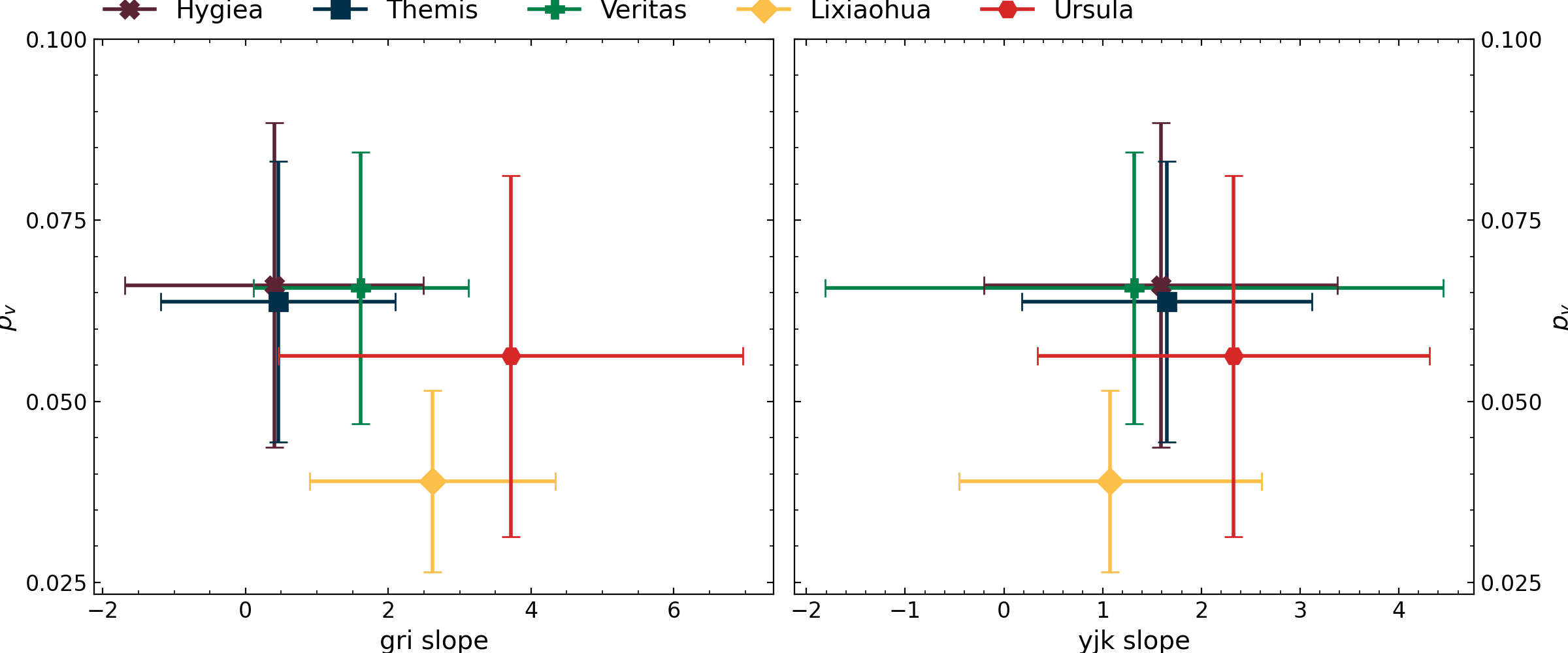}
                \caption{Weighted average visible slope (\textit{left}) and NIR slope (\textit{right}) vs. albedo for five outer-belt asteroid \textbf{families}.}
                \label{fig:catalogs}
        \end{figure*}

        To obtain asteroid reflectance spectra, we divided the object spectrum by the spectra of solar analogs. We analyzed the spectra of the solar analog stars to detect small differences in color introduced during the observations, caused for example by inconsistent centering of the star in the slit. These differences could propagate into the reflectance spectrum of the target through the reduction process. The first step was to apply an atmospheric extinction correction.      This effect is dependent on the airmass and the wavelength, with shorter wavelengths  and higher air masses experiencing greater extinction. To minimize the extinction effect from the difference in airmass between the stars and the target, we applied a correction to the spectra of the object and the stars. In the absence of extinction coefficients for Cerro Pach\'on, we used the mean extinction coefficients for La Silla, because this observatory is located relatively close and at similar altitude to Cerro Pach\'on. A study of the variation of extinction coefficients from different sites suggested it is mostly influenced by the altitude of the site. For each night, we divided all of the spectra of the solar analogs by one used as reference, typically the one at lower airmass. Ideally, the result of this division should be a slopeless flat spectrum, with a normalized flux value of around 1. The star observations that appeared to present outliers (i.e.,  an uncommon shape relative to the others and/or spectral inclination variation above 5$\sigma$)  were discarded and the asteroid reflectance spectra were obtained using the remaining stars. Finally, all reflectance spectra were normalized to 1 at 0.55$\mu$m.
        
        The data reduction was made by combining scientific Python libraries \citep{numpy, scipy,astropy:2013,astropy:2018, matplotlib} with IRAF\footnote{IRAF is distributed by the National Optical Astronomy Observatories, which are operated by the Association of Universities for Research in Astronomy, Inc., under cooperative agreement with the National Science Foundation.} tasks, called through the PyRAF\footnote{PyRAF is a product of the Space Telescope Science Institute, which is operated by AURA for NASA.} library.

        \section{Analysis}
        \label{sec:ana}
        
        In addition to the 29 spectra observed, we collected all available spectra in the literature of members of Themis, Hygiea, Veritas, Ursula, and Lixiaohua families \citep{1996PDSS..168.....X, 2004PDSS....1.....B,2006PDSS...45.....V,  2007PDSS...51.....L, de2020spectroscopic}. In particular, the spectroscopic analysis of the Lixiaohua was recently presented in \cite{de2020spectroscopic}, in which we used the same methodology described in this section. Therefore, the individual analyses of each spectrum of the Lixiaohua family are not presented in this paper. In addition to the spectroscopic characterization, we include the albedo, diameters, visible and NIR colors, and rotational periods of each family. Table \ref{tab1} includes the orbital parameters of the largest member of each family, the family age from \cite{2013A&A...551A.117B}, the number of members with observed spectra (both observed in this work and taken from the literature), members with SDSS colors \citep{carvano2010, Hasselmann2011}, NIR MOVIS colors \cite{Popescu2016}, WISE albedos and diameters \cite{mainzer2016}, and rotational properties with U>2 \citep{2009Icar..202..134W}.

        \subsection{Spectroscopic analysis}
        The entire analysis of the spectroscopic data was performed using the CANA package \citep{cana}\footnote{CANA code is available though https://cana.readthedocs.io}. In order to \textbf{analize the spectra}, we performed a taxonomical classification, and we also computed (whenever the wavelength coverage allowed it) four different parameters. These parameters are: visible slope (visible - red-mode), NIR slope (NIR), hydration feature (visible - red-mode), and turn-off point (visible - blue-mode).
        
        The visible slope was measured using the angular coefficient in a linear fit to the spectrum in the wavelength range of 0.5-0.9 $\mu m$, normalized to 0.55 $\mu m$. The spectral slopes are  expressed in units of $\%/1000$\AA. The associated error is the sum of the systematic error introduced by the solar analogs division, and the error of the fit estimated by a Monte-Carlo method with 1000 iterations, where the spectra reflectance was resampled considering the spectrum S/N. Although the results presented are the quadratic sum of the described error sources, the systematic error is the strongly dominating term. 
        
        We looked for the presence of hydrated minerals through the identification and characterization of an absorption band centered at around 0.7 $\mu m$. For this task we used the wavelength range of 0.55-0.88 $\mu m$. The spectral continuum was estimated by a linear fit within the      0.55-0.57 and 0.84-0.88  $\mu m$ intervals. Thereafter, we removed the continuum from the spectrum and fit a fourth-order spline in the 0.55-0.88 $\mu m$ range. We identified the hydration band if the minimum reflectance of the fit was close to 0.7 $\mu m$, at a depth greater than $1 \%$ and of  $3\sigma$ of the spectrum S/N.
        
        A common feature in primitive asteroids is a turn-off in the slope at approximately 0.5 $\mu m$. We characterized the presence of this feature using the data available in the 0.4-0.7 $\mu m$ range, applying the same methodology as described in \cite{2018Icar..311...35D}.

        \begin{figure*}[t!]
        \centering
        \includegraphics[scale=0.55]{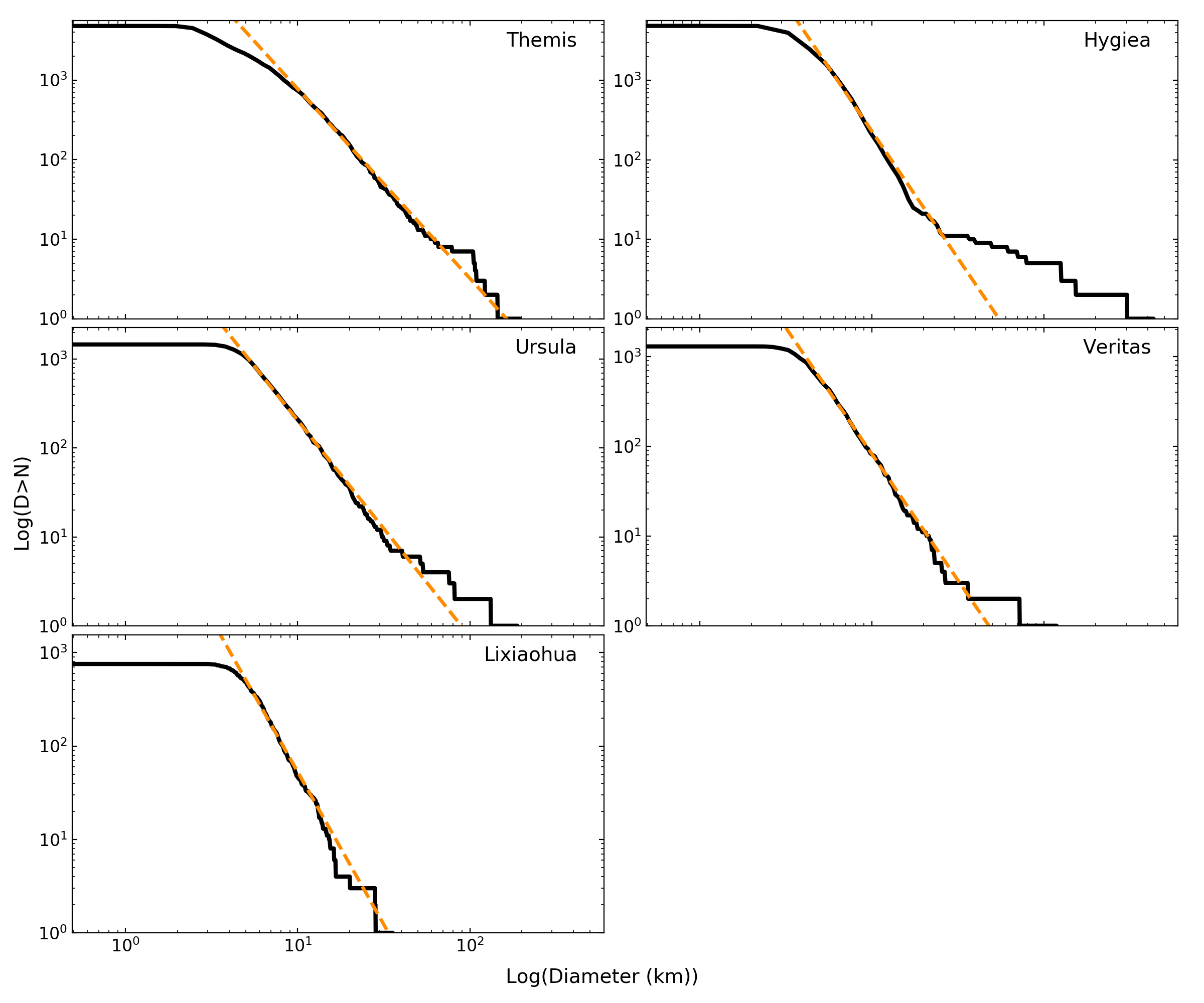}
        \caption{SFD for the five outer-belt asteroid family. The orange line represents the power-law fit that was used to include the contribution of small-sized members into the calculation of the family's parent body size.}
        \label{fig:teste}
\end{figure*}
        
        Taxonomic classification of the asteroids was carried out by minimization of the chi-squared between the spectrum of a given object  and the templates from the \cite{2009Icar..202..160D} taxonomic system. This procedure was performed using the combined spectra with data from all wavelength ranges available. If the spectrum was sampled only at visible wavelengths, we adjusted the class to fit the \cite{bustaxonomy} scheme. The results for the spectroscopic analysis are shown in Table \ref{tab:results}.

        \subsection{Public photometric catalogs}
        
        In addition to the spectroscopic data, we collected data of the families available from public databases. Visible colors were extracted from the SDSS  \citep{carvano2010, Hasselmann2011}, NIR color from the MOVIS catalog \citep{Popescu2016}, and geometric albedos from the NEOWISE survey \citep{mainzer2016}.
        
        We calculated the visible and NIR spectral slopes from the data available in these databases such that they are comparable with the ones derived from the spectroscopic data. We estimated the angular coefficient from a linear fit using the reflectances on g, r, and i SDSS filters, normalized at 0.55 $\mu m$ for the visible slope; and using the Y, H, and K VISTA filters, normalized in 1.2 $\mu m$ for the NIR slope. The criteria for the choices of filters was based on selecting the highest number of objects that were observed in at least three filters and covering almost the same wavelength range as the spectroscopic data ($0.5-0.9~\mu m$ and $1.0-2.3~\mu m$).  To estimate the uncertainties, we created 1000 clones of each observation by drawing random values for the reflectance in each filter using normal distributions with means equal to the listed reflectance value and variances equal to the listed uncertainties. The resultant spectral slope distribution was then fitted with a Gaussian curve, the mean and variance of which were then adopted as the nominal value for the spectral slope and its uncertainty, respectively, expressed in units of $\%/1000$\AA. Finally, the mean visible and NIR slopes of each family were then estimated by a weighted average, using the slope uncertainties as the weights. A weighted average, with the uncertainties provided by the NEOWISE catalog, was used to estimate the mean albedo for each family. The results of this analysis are shown in Fig. \ref{fig:catalogs} and Table \ref{table:results}.

        \begin{figure*}[t!]
                \centering
                \includegraphics[scale=0.55]{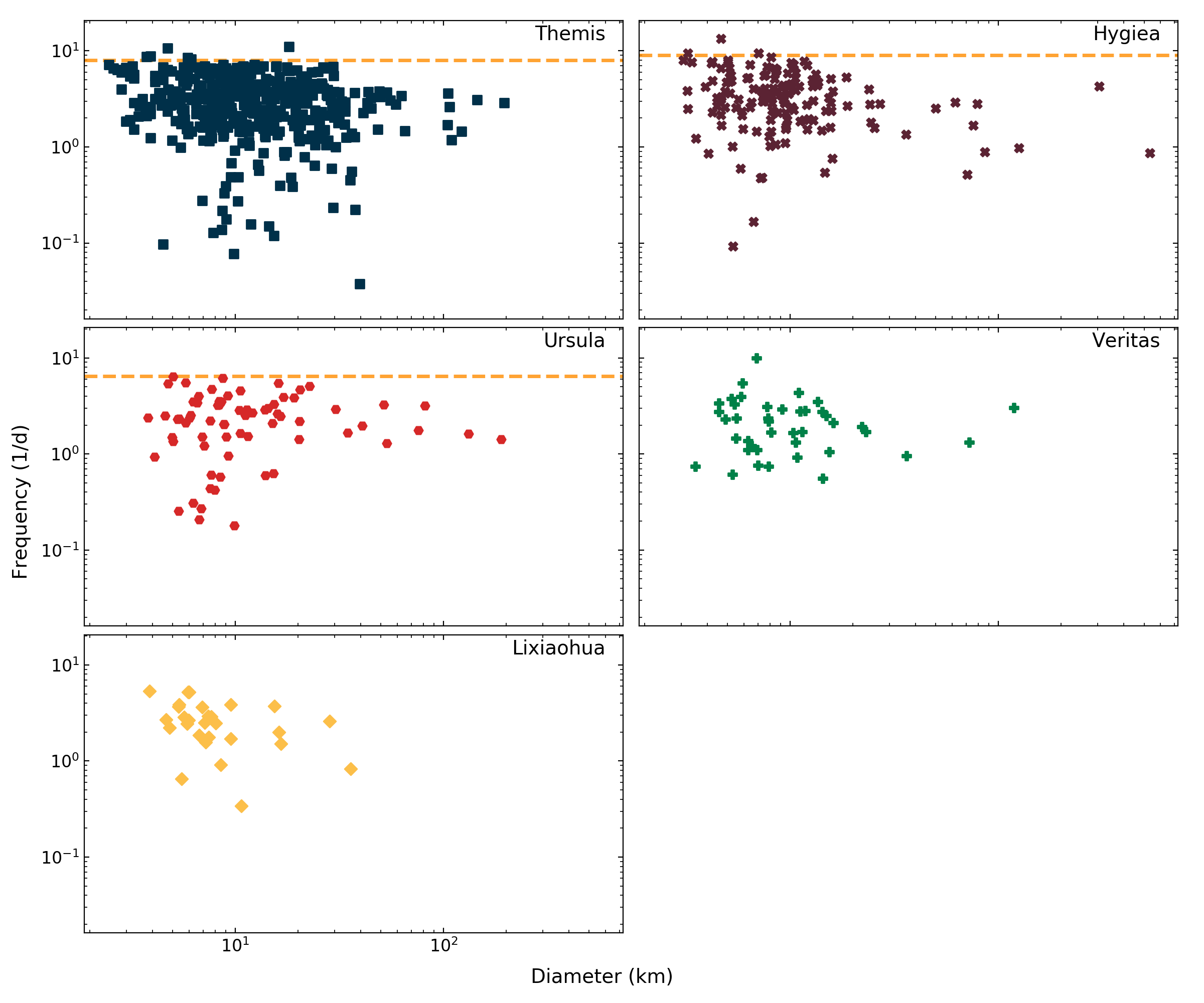}
                \caption{Rotation frequency vs. diameter for each family. The orange line delimits the estimated critical period. For Veritas and Lixiaohua, the small number of determined periods prevents the determination of this value.}
                \label{fig:rotation}
        \end{figure*}

        \subsection{Parent body sizes}
        
        We estimated the parent body diameter (PB) for  each family based on the work of \cite{2008MNRAS.390..715B}. To build the size--frequency distribution (SFD), we obtained diameters of the family members from the WISE survey \citep{mainzer2016}. Objects with a geometric albedo $p_v>0.12$ were removed in order to consider only objects that could have a primitive surface and minimize the presence of interlopers. For the objects that did not have a determined diameter, we estimated it using the following equation:
        
        \begin{equation}
                D = \frac{1329\times10^{-0.2H}}{\sqrt{p_{fam}}}
        ,\end{equation}
        where H is the absolute magnitude of the object and $p_{fam}$ is the mean geometric albedo of the respective family. Figure \ref{fig:teste} shows the SFD for each family, where it is possible to observe that there is an inflection point in the diameter distribution towards the smaller objects. We consider, as an approximation, that this inflection is related to an observational bias and use these points as the cutoff in diameter ($D_c$) to define the sample completeness. The volume of the PB ($V_{PB}$) can be estimated using:

        \begin{equation}
                V_{PB} = V_{c} + \int_{0}^{D_{c}} (\alpha \log_{10}(D)+\beta)~dD
        ,\end{equation}
        where $V_c$ is the the sum of volumes of the objects with $D>D_c$. The second term is the integral of a power-law fitted considering $D>D_c$, where $\alpha$ and $\beta$ are the coefficients of the fit. The integral limits are set from $0$ to $D_c$ in order to add the contribution of the small-sized members. Finally, the diameter of PB  ($D_{PB}$) is estimated assuming a spheroidal shape:
        \begin{equation}
                D_{PB}= \sqrt[3]{\frac{6}{\pi}V_{PB}}
        .\end{equation}
         
         Figure \ref{fig:teste} shows that, with the exception of Hygiea, the SFDs for all famailies are well approximated by a power law in the regime of $D>D_c$. In the case of Hygiea, we also defined an upper limit in diameter to constrain the region where the power law was fitted.
         
         Although we use the power-law fitting to minimize the errors on the parental body size derived from the observational bias on smaller members, there are several sources of error in this estimate. For example, the individual errors in the diameter of the family members available in the WISE catalog are not considered here. Additionally, family members can be removed from the region, for example through Yarkovsky-driven evolution or chaotic diffusion by crossing resonances with giant planets. The process followed to identify families considers a cutoff in velocity, which typically leaves a halo of potential family members with similar surface properties, but with slightly distinct orbital parameters that are not considered as belonging to the families \citep{2015aste.book..297N}. Finally, even though we defined a range in geometric albedo to minimize the presence of interlopers, there should still be objects that are not true members of the family but that occupy the same region and may have been added in the SFD. The procedure of selecting only the family members with albedo $p_v>0.12$ revealed that 1 to 12\% of the family members can be outliers.

        \subsection{Density}
        
        %Also note that the density of a body depends both on the bulk density of the material and the total porosity (macro+micro) in the body. Bulk density is related to the composition (but should vary within a parent body as function of depth); The porosity, on the other hand, is linked mostly to the collisional history of the body.  Since all bodies will have some porosity, the calculated porosity will necessarily be smaller than the bulk density of the material that composes the body. If you use the smallest period in the family, you would get a lower limit for the observed density in the family; you could then discard materials with bulk densitiy smaller than that limit. 44
        
        The analysis of asteroid rotational properties performed by \cite{2000Icar..148...12P} showed that objects larger than 0.15 km do not rotate faster than 11 revolutions per day. This spin barrier is interpreted by means of a  rubble-pile structure  model, in which rotation with periods exceeding the critical value would cause asteroid breakup \citep{2007Icar..190..250P}. The smaller objects with faster rotation are thought to be monolithic collisional fragments with stronger cohesion forces. 
        
        In a cohesionless rubble pile, the critical value for the spin barrier is dependent on bulk density (Eq. \ref{eq:density}). The bulk density of a body depends  on both the grain density of the material and the total porosity in the body. The grain density is related to the composition, while the porosity is linked mostly to the collisional history of the body. Asteroid families with distinct formation histories, composition, and differentiation level should present differences in densities, which would be reflected by distinct spin barriers. 
        
        We performed an estimate of the density of the family members based on their rotational properties. This analysis considers two a priori hypotheses: 1) all objects with diameter within the 4-20km range in a family that are close to the spin barrier have the same density, and 2) the original spin distribution reached the spin barrier, or the family has existed for a sufficiently long time for YORP effects to increase the spin up to the spin barrier. Considering this scenario and assuming that the structure of objects with diameters in the 4-20 km range can be approximated by a cohesionless rubble pile \citep{2000Icar..148...12P}, we can use the following equation to infer these properties:

        \begin{equation}
        \rho \approx \frac{(1+\Delta m)}{(P_{crit}/3.3)^{2}}
        \label{eq:density}
        ,\end{equation}
        where $\rho$ is the bulk density, $P_{crit}$ is the critical value for the rotation period (in hours), and $\Delta m$ is the amplitude of the rotational light curve. For this analysis, we use a spherical shape approximation, where $\Delta m=0$. 
        
        To determine the critical value for each family, we searched for cataloged rotational periods of family members in the Asteroid Light curve  DataBase (LCDB), Updated 2019 August 14 \citep{2009Icar..202..134W}, with the U parameter of the light curve equal to or higher than  two. The U parameter is defined based on the quality and coverage of light curve used to estimate the rotational period, a value of 2 is the recommended value for statistical analysis and indicates that the light curve was not completely observed and that the period uncertainty can reach up to 30\%. The number of objects per family that satisfies this criteria is shown in Table \ref{tab1}. Figure \ref{fig:rotation} shows the frequency versus diameter plot for each family with the respective critical period value.
        
        We note that the number of cataloged rotational periods per family can be too small for the determination of a spin-barrier. This is seen especially in the case of the Lixiaohua and Veritas families. Additionally, these families possess the two youngest estimated ages in our sample (Table \ref{tab1}), and their spin distribution could be less affected by the YORP effect and collisional evolution, so much so that they do not yet present a spin barrier. We note that the estimated densities are the result of a tentative approach that will become more robust as we increase the number of determined rotation periods.
        
        %%%%%%%%%%%%%%%%%%%%%%
        \section{Results}
        \label{sec:res}
        
        \begin{table*}[h!]
        \footnotesize
        \centering
        \setlength\tabcolsep{2.2pt}
        \begin{tabular}{lcccccccccccc}
                \hline
                Asteroid & Vis. Slope & Vis. Slope &IR Slope &Turn-off & Hydrated & 3-micron & MBC &  Albedo & PB & PB & LM/PB & Density \\
                Family & spec. & SDSS & MOVIS & members & members & feature & members &   & TW  & Durda07  & TW  &\\
                 & (\%/1000\AA)& (\%/1000\AA) & (\%/1000\AA) & Nf (Ns) & Nf (Ns) & (shape) & (number) & & (km) & (km) & & \\
                \hline
                Themis    & $-0.2\pm1.6$ & $0.5\pm1.6$ & $1.7\pm1.5$& 11 (21) & 8 (59) & rounded        & 3 & $0.066\pm0.033$ & 286 & 451 & 0.68 & 1.2 \\
                Hygiea    & $0.3\pm3.7$ & $0.4\pm 2.1$& $1.6\pm1.8$ & 10 (13) & 3 (25)  & ceres/rounded  & 0 & $0.068\pm0.042$ & 492 & 442 & 0.88 & 1.5  \\
                Veritas   &$1.0 \pm 1.3$ & $1.6\pm 1.5$& $1.3\pm3.1$ & 14 (15) & 13 (17) & -              & 0 & $0.068\pm0.062$ & 135 & 177 & 0.88 & - \\
                Ursula    &$2.8\pm1.5$ & $3.7\pm 3.3$ & $2.3\pm2.0$ & 4 (4) & 0 (7) & rounded        & 0 & $0.058\pm0.063$ & 226& 280 & 0.83 & 0.8  \\
                Lixiaohua & $3.0\pm3.2$ & $2.8 \pm 1.7$ & $1.1\pm1.5$&  5 (30) & 0 (35) & -              & 2 & $0.043\pm0.081$ & 69 & 220  & 0.51 & - \\
                \hline
        \end{tabular}
        \caption{Average physical properties of outer-belt families. We present values for parent-body diameters (PB) obtained in this work (TW) and by \cite{2007Icar..186..498D} (Durda07) for comparison. The ratio between the diameter of the families' largest member and that of the parent body (LM/PB) are shown only for diameters obtained in this work. Nf represents the number of objects that present the feature, while Ns is the number of spectra for which the wavelength was appropriate to perform the feature analysis.}
    \label{table:results}

        \end{table*}

        \subsection{Themis}
        
         We increased the number of members with observed spectra to reach a total of 59 family members by adding family data available in the literature \citep{1999A&AS..134..463F, 2005Icar..174...54M, 2011Icar..213..538Z, 2012A&A...537A..73L, 2012Icar..218..196D, 2016Icar..264...62K, 2016Icar..269....1F, 2016Icar..269...62L, 2016A&A...586A..15M}. The mean visible slope was \textbf{estimated at} $-0.17\pm1.58\%/1000$\AA, and 13\% of the sample showed the 0.7$\mu m$ feature. The majority of the objects were classified as belonging to C-class, with a significant fraction of B-types, hence the negative mean slope. This value is consistent, considering the errors, with the $0.45\pm1.64\%/1000$\AA~estimated using SDSS data for 563 objects. There were 21 out of the 59 spectra with appropriate wavelength coverage for the Turn-off analysis, 11 of which showed this  feature. The critical wavelength for the Turn-off change can vary significantly from  0.49-0.67 $\mu m$. The MOVIS data provided a mean NIR slope of $1.65\pm1.47\%/1000$\AA, slightly higher than the $0.12\pm0.07\%/1000$\AA\  obtained using spectroscopic data by \cite{2011Icar..213..538Z}. The latter work used the wavelength range of 1.6-2.3 $\mu$m to measure the slope, which is distinct from the 1.0-2.15 $\mu$m provided by the MOVIS data, which was used in this work. Additionally, these latter authors found that many Themis family members can show a concave positive curvature in the spectra. Both factors could be influencing the slope determination from the photometric data. 
        
        The Themis parent body size was estimated to be approximately 286 km, in agreement with the value obtained by \cite{2013A&A...551A.117B}, which was calculated in a similar manner. However, this value is considerably smaller than those obtained using different techniques by \cite{2007Icar..186..498D}  and \cite{1999Icar..141...65T}, of 451 km and 369 km respectively. Nonetheless, the estimated values of PB provide a range of $\slfrac{LM}{PB}$ (largest member diameter divided by the parent body size)  from $0.51-0.68$, which is compatible with a catastrophic event being the origin for the family.
        
        The analysis of the rotational properties of 443 Themis family members presented the most well-defined spin barrier among the studied families. The critical period at $\sim3$h provides an estimate of the bulk density of $1.2$ g/cm$^3$. The bulk density of (24) Themis was estimated at $1.81\pm0.67$ g/cm$^3$ by \cite{2012P&SS...73...98C}, while \cite{2007Icar..187..482D} studied the binary asteroid (90) Antiope, a member of the Themis family, and found a density of $1.25\pm0.05$ g/cm$^3$. Both values are in agreement with the estimate in this work, accounting for the error bars. Another binary member of the family, (379) Huenna, was studied by \cite{2008Icar..195..295M}, who found a bulk density of $0.9\pm0.1$ g/cm$^3$, slightly under the value estimated by the spin-barrier method. However, the diameter used to estimate (379) Huenna bulk density was higher than the one provided by \cite{mainzer2016}, and a smaller diameter would return a higher value for the density. Additionally, it is of note that larger objects ($D>30$ km) can follow a different regime in terms of cohesive forces and density, as proposed by \cite{2000Icar..148...12P}.

        \subsection{Hygiea}
        The Hygiea family showed very similar properties to the Themis family at the studied wavelengths, with a mean slope of $0.30\pm3.68\%/1000$\AA~(spectroscopic) and $0.40\pm 2.09\%/1000$\AA~(photometric) in visible wavelengths, and $1.07\pm1.53~\%/1000$\AA~ in the NIR. However, there is a higher diversity in spectral slope than in Themis. Most of the targets were also classified as C-class, with a significant B-type contribution. Approximately $15\%$ of the family members present the 0.7 $\mu m$ hydration band, and 12 out of 17 objects show the turn-off feature.

        Asteroid (10) Hygiea is the largest object in the outer-belt region, and the fourth largest in the main belt. The diameter cataloged by \cite{mainzer2016} for (10) Hygiea is of $\sim533$ km, considerably higher than the previous estimate from \cite{1994PDSS..101...96T} of $\sim407$ km. A recent work of \cite{2019NatAs.tmp..477V} using a direct imaging technique estimated an equivalent diameter of $\sim434$ km. We adopted this latter value for the object and to construct the family SFD. The parental body size was estimated at 486 km, with $LM/PB\sim0.89$.  \cite{2007Icar..186..498D} estimated the family PB to be 440 km, in good agreement considering that these latter authors used a different technique and a smaller diameter for (10) Hygiea. Figure \ref{fig:teste} reveals an excess of large asteroids which is characteristic of a shattering event, where most of the family mass is distributed among the largest members. These values are influenced by the presence of other large asteroids besides (10) Hygiea, such as (52) Europa and (159) Aemilia, with respective diameters of $\sim303$ km and $\sim125$ km. Both of these objects were identified as Hygiea family members by \cite{2015aste.book..297N}, but not in \cite{2019NatAs.tmp..477V} and \cite{2013MNRAS.431.3557C}. The exclusion of these objects from the family would produce a smaller parent body, where the second-largest body would be (1599) Giomus with a diameter of 52 km. The presence of (52) and (159) can significantly change the interpretation of the collisional event that formed the family and of the nature of the family parent body. 
        
        A rotation period has been derived for 149 members of the Hygiea family. The spin barrier is defined at $\sim2.7$h, which provides a density estimate of $1.5$ g/cm$^3$. In \cite{2019NatAs.tmp..477V}, the authors estimated a bulk density of $1.944$ g/cm$^3$ for (10) Hygiea, which is in reasonable agreement with the $2.19\pm0.42$ g/cm$^3$ calculated by \cite{2012P&SS...73...98C}. These values are considerably higher than our estimates. However, (10) Hygiea is considerably larger than the regime where we use the cohesionless approximation (4-20 km). Also, considering that this object could be a differentiated body, and could have different layers of material and levels of compaction, a higher value of this object density is plausible and not contradictory to the possibility that smaller objects could have lower densities.  \cite{2012P&SS...73...98C} also estimated a density of $1.52\pm0.39$ g/cm$^3$ for (52) Europa, which is in agreement with the value obtained in this work.
        
                \begin{figure}[h!]
            \centering
            \includegraphics[scale=0.6]{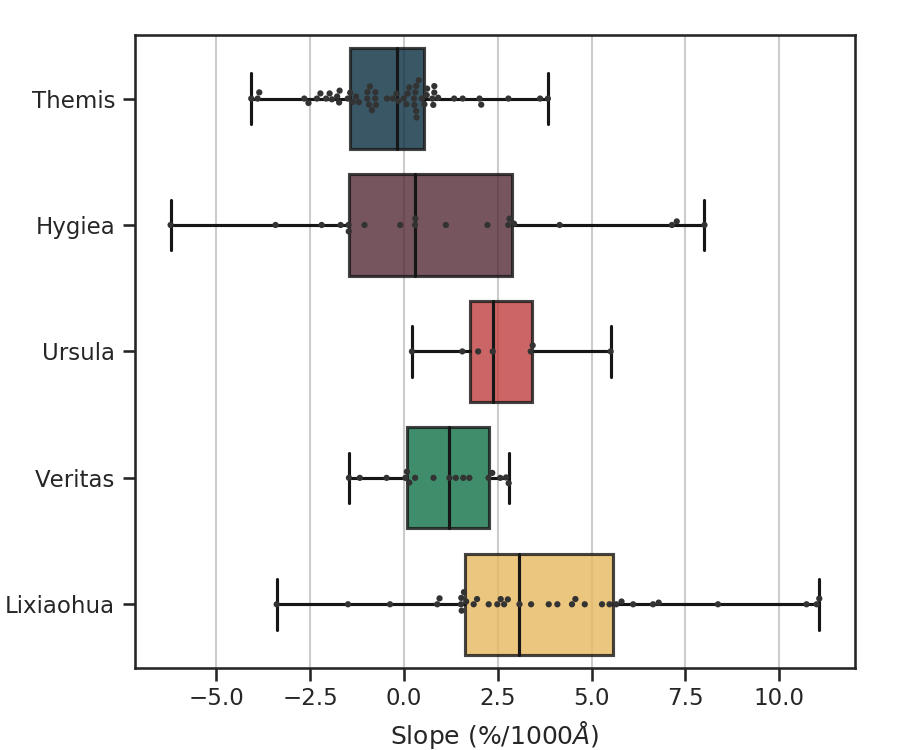}
            \caption{Whisker plot of the visible slope distribution in each family.  The boxes represent the average and standard deviation of the slope distribution, while the solid black line inside of the box represents the median. The black dots represent single slope measurements of family members, with the error bars showing the maximum and minimum values.}
            \label{fig:slopes}
        \end{figure}
        
        \subsection{Veritas}
        The Veritas family showed the highest spectral homogeneity among its members, with a mean visible slope of $1.00\pm1.30\%/1000$\AA~ (spectroscopic) and $1.61\pm1.50\%/1000$\AA~ (photometric) and a high content of hydrated members reaching 77\% of the sample. Of the 16 objects analyzed in this family, 14  show the Turn-off feature. The majority of the objects' spectra were classified in the C-complex, especially as Ch-type, which reflects the high level of hydrated members in the family. This result is in agreement with that of \cite{2005Icar..174...54M}. The mean NIR slope was estimated at $1.31\pm3.12\%/1000$\AA, higher than the $-0.04\pm0.03\%/1000$\AA\  obtained using spectroscopic data obtained by \cite{2011Icar..213..538Z}. These latter authors noted that, in contrast to the Themis family, the NIR spectra for the Veritas family show a concave negative curvature. As in the Themis family case, the wavelength range used to determine the slope and the curvature in the spectra could be influencing the different values obtained in this work.

        The Veritas parental body size was estimated at 136 km, which is slightly smaller than the 177 km obtained by \cite{2007Icar..186..498D}. Both values are in agreement with those of \cite{2013A&A...551A.117B}, who estimated the Veritas PB at 100-177 km. The small number of cataloged rotational periods for Veritas members prevents us from obtaining an estimate of the family spin barrier and density limit.

        \subsection{Ursula}
        
        The Ursula family present a spectroscopic mean slope of $2.77\pm 1.46\%/1000$\AA, compatible with the $3.71\pm 3.25\%/1000$\AA~ estimated using SDSS data. Based on NIR colors, Ursula showed the reddest mean slope among the studied families. None of the objects exhibited the 0.7 $\mu m$ feature, indicating that the family might have a low hydrated
mineral content. Although the spectra for (375) Ursula obtained by \cite{2004PDSS....1.....B} and \cite{2006PDSS...45.....V} presented similar spectral slopes, there is a significant difference in the shape of the spectra, and the data from the latter work might suggest a hydration feature.
        
        We estimated the parent body diameter at 226 km, which is slightly larger than the 203 km estimated by \cite{2013A&A...551A.117B} and smaller than the 280 km measured by \cite{2007Icar..186..498D}. The rotational distribution suggests a spin barrier with a critical rotational period of $\sim3.6$h, which provides an estimate for the density for the Ursula family of $\sim0.8$ g/cm$^3$ .

        \subsection{Lixiaohua}
        
        The results for the spectroscopic analysis of the Lixiaohua family are described in \cite{de2020spectroscopic}, where we found a mean slope of $3.03\pm3.22\%/1000$\AA~ (spectroscopic) and $2.79 \pm 1.72\%/1000$\AA~ (spectrophotometric) in visible wavelengths, and of $1.07\pm1.53\%/1000$\AA~ in the NIR. The Lixiaohua is the darkest of the studied families, with mean albedo of $0.043\pm0.081$ (Fig. \ref{fig:catalogs}).
        
        In this work, we extended the analysis by calculating the parent body size, estimated at $PB\sim69$ km and the ratio $\slfrac{LM}{PB} \sim 0.51$, which are consistent with a catastrophic disruption event. This PB value is significantly smaller than the one estimated by \cite{2007Icar..186..498D}, of $\sim 220$ km, who also provide $\slfrac{LM}{PB} \sim 0.16$. The latter value suggests that the majority of the parent-body mass was scattered or lies in the smaller undiscovered family members. There are only 28 Lixiaohua members with determined rotational periods. The analysis of Fig. \ref{eq:density} suggests that the family spin barrier might be considerably low. However, more data are needed to obtain any conclusive density estimate.

        \section{Discussion}
        \label{sec:dis}
        
        The analysis of the outer-belt families Themis, Hygiea, Veritas, Ursula, and Lixiaohua reveals a diversity of physical properties. In Table \ref{table:results}, we present the results for the spectroscopic properties, parent-body sizes, densities, and include information that contributes to this discussion:  the 3 $\mu$m band information extracted from the literature \citep{2010Natur.464.1320C, 2012Icar..219..641T, 2019JGRE..124.1393R} and the presence of  main-belt comets \citep{2018AJ....155...96H}. In particular, the presence of hydrated and ice-bearing objects in a single family can be considered an indication of partial differentiation of the parental body, as proposed for the Themis family. \cite{2010GeoRL..3710202C} and \cite{2016A&A...586A..15M} suggested that these two materials could be representatives of different layers in the internal structure of the family's parent body. Hydrated minerals, which are typically formed by the interaction of liquid water with silicates such as olivine, would have formed in the inner layers that experienced higher temperature and pressure conditions, while the water ice could remain nearly unaltered in the external layers. In contrast, an undifferentiated body should present members with only hydrated minerals or water ice, but not both.
        
        Figure \ref{fig:slopes} shows the median and mean slope of each family with associated variance and minimum and maximum values, while Figure \ref{fig:catalogs} shows the albedo visible and NIR slopes obtained from public catalogs. Themis, Hygiea, and Veritas families show neutral visible slopes compatible with C-type asteroids, with a higher variability in the Hygiea family. In contrast, Ursula and Lixiaohua show slightly reddish slopes, related to P-type objects. 
        
        We identified different hydrated fractions among the families. The two large C-type families Themis and Hygiea show a similar percentage of hydrated members at $13-15\%$, while the smaller Veritas family shows a higher fraction of hydrated members, reaching $87\%$. The Veritas family also shows the highest fraction of members showing the turn-off feature. This feature in C-type objects is attributed to ferric oxide intervalence charge transfer transition \cite{1985Icar...63..183F,1995Icar..115..217V}. \cite{2019GeoRL..4614307H} proposed the absence of the feature could be related to a higher degree of space weathering. This result could explain why Themis is the oldest family in our sample  and shows a lower fraction of turn-off members, in contrast to Veritas,  which is the youngest and might show younger surfaces. However, it is worth noting that Themis and Hygiea are considerably closer in age than each is to Veritas. 
        
        Ursula and Lixiaohua show no unambiguous indication of hydrated members. One difference between the spectra of these families is seen in the fraction of members that exhibit  the Turn-off feature. However, the analysis of the Ursula family is based on only four objects and more data are needed to make a proper comparison of this feature between these families.
        
        The sample of families covers a wide range of parental body sizes, from 69 to 492 km. The approach used by \cite{2007Icar..186..498D} compared the SFD of the families with ones simulated using a combination of smooth-particle hydrodynamics (SPH) and N-body codes. Their results provide estimates typically larger than those calculated in this work, with the exception of the Hygiea family. The major differences are found in the case of Themis and Lixiaohua families. In both cases, the higher parental body size suggests that most of the mass is located in the small-sized members, or was detached from the family through dynamical evolution, or lost in the dust produced by the collision. While that could be the case for Themis, it is unlikely that the largest fraction of the mass from the Lixiaohua progenitor was lost, and a smaller parent body size than the one proposed by \cite{2007Icar..186..498D} is more likely. 
         
        \cite{2012P&SS...73...98C} estimated an average bulk density of $1.33\pm0.58$ g/cm$^3$ for C-type objects, in agreement with the values we estimated for the Themis and Hygiea families. \cite{2017P&SS..147....1C} used a similar methodology to the one described in this work and found a lower limit for the density of C-types of $0.90\pm0.05$ g/cm$^3$. A possible interpretation for this scenario is that there could exist C-type asteroids with lower densities than the members of Themis and Hygiea, which could be explained by variations in grain density (composition) and porosity. For the  Ursula family, the low value for density is compatible with the P-/D-types and cometary nuclei \citep{2012P&SS...73...98C}, which is in agreement with the family spectroscopic properties. 
         
        The largest members of Hygiea, Themis, and Ursula have been observed in the 3.0 $\mu$m region. The Themis family members (24) Themis and (90) Antiope both show a "rounded" feature associated with the presence of water-ice on their surfaces \citep{2010Natur.464.1320C, 2010Natur.464.1322R,2012Icar..219..641T}. \cite{2019PASJ...71....1U} presented the spectrum of (24) Themis obtained by the Akari space telescope, where they identified two features: one at 2.7 $\mu$m associated with hydrated minerals, and another at around 3.071 $\mu$m related to water ice. Another indication of water ice as a constituent of the family composition is the presence of three MBCs, in which the cometary activity is suggested to be sublimation-driven \citep{2018AJ....155...96H}. As proposed by \cite{2010GeoRL..3710202C} and \cite{2016A&A...586A..15M}, a possible explanation for the spectral diversity of Themis, which includes hydrated members, MBCs, and surface ice on two family members, is that the Themis parent body could be partially differentiated.
        
        The spectrum of (375) Ursula was observed by \cite{2003M&PS...38.1383R}. At that time the absorption feature was considered to be similar to the one in (1) Ceres. However, as noted by \cite{2019JGRE..124.1393R}, the band would now be classified as similar to the one in (24) Themis, also suggesting the presence of water ice in the family. In contrast with Themis, the Ursula family shows little to no sign of hydrated members. The spectroscopic properties and the low density of the Ursula family suggest that the parent body was formed in a similar region to the P-/D-type asteroids and possible cometary nuclei, beyond the snow line.
                
        In the case of  the Hygiea family, the analysis by \cite{2012Icar..219..641T} of (10) Hygiea showed an absorption feature classified as similar to the one in (1) Ceres \citep{2012Icar..219..641T}. However,  \cite{2019JGRE..124.1393R} recently acquired seven additional spectra for the object, three of which were described as ``rounded' or ``Themis-like'', three as ``Ceres-like'', and one as hydrated (``Sharp'' or ``Pallas-like''). The Ceres-like absorption feature is interpreted as being due to the presence of ammoniated minerals, while the Sharp or Pallas-like feature is related to the presence of hydrated minerals. \cite{2019PASJ...71....1U} identified  two significant  features  at  2.72  and  3.08 $\mu m$ in the spectrum of (10) Hygiea and a small 3.1 $\mu m$ feature as well as a weaker feature at 2.7 $\mu m$ in the spectrum of (52) Europa. These features are related with hydrated minerals and water ice, respectively. \cite{2019JGRE..124.1393R} also acquired four spectra of (52) Europa, the second-largest member of the Hygiea family, and all of these presented rounded or Themis-like features, attributed to water-ice. However, \cite{2012Icar..219..641T} suggested that (52) Europa shows a distinct band shape classified as Europa-like, which could lead to a distinct compositional interpretation. The variability in IR spectra for (10) Hygiea could be attributed to a heterogeneous surface. However, \cite{2001Icar..152..117M} and \cite{2019NatAs.tmp..477V}, using different techniques, did not find evidence of strong surface inhomogeneity for this object. As suggested by \cite{2019JGRE..124.1393R}, additional data are needed to fully characterize the surface of (10) Hygiea.
        
        \cite{2013MNRAS.431.3557C} and \cite{2014Icar..243..429R} suggested that (10) Hygiea could be an interloper of the family. However, the object is consistent with the family members at the wavelength regions analyzed in this work. Additional investigations at different wavelengths are necessary to confirm or reject this hypothesis. The large diameter of the family's parent body and the similarity of surface properties between (10) Hygiea and the differentiated body (1) Ceres \citep{2019JGRE..124.1393R} suggest that the parent body and/or (10) Hygiea have also undergone differentiation.   
                        
         On the other hand, Veritas and Lixiaohua do not seem to be differentiated. As discussed in \cite{de2020spectroscopic}, the Lixiaohua family shows evidence of water ice through the presence of two MBCs among its members and no trace of aqueous alteration. In contrast, the Veritas family shows a large fraction of hydrated members, which indicates compositional homogeneity. \cite{2011Icar..211..535M} proposed that the asteroid (490) Veritas, the largest member of the homonymous family, could be an interloper in the family. However, the spectroscopic properties of the asteroid are consistent with a highly homogeneous family.
         
         The sizes of the parent bodies of both the Lixiaohua and Veritas families are below 150 km, which is considerably smaller than the parent bodies of Themis and Hygiea. Combined with their spectroscopic properties, this suggests that these families did not originate from the break-up of differentiated bodies. The differences in spectral properties between these two families suggest that they were formed at different places and/or times in the early Solar System, where the Veritas parent body might have formed sooner and was more exposed to radiogenic heating.

\section{Conclusions}
    We present new spectroscopic data for the outer-belt families Themis, Hygiea, and Veritas, and include data available in the literature for the Ursula and Lixiaohua families. Our analyses reveal the diversity of surface properties among the families, where the Themis, Hygiea, and Veritas families present neutral mean slopes in the visible and NIR wavelengths. In contrast, Ursula and Lixiaohua families present reddish mean slopes. The analysis of the albedo reveals that the Lixiaohua is also the darkest among these families. The number of hydrated members varies significantly:   77\% of Veritas members show the 0.7 $\mu$m feature, while Lixiaohua and Ursula show no evidence of hydration. 
    
    In addition to the spectroscopic analysis we also estimated the parental body size of each family and the density based on their rotational properties. The parental body sizes of Themis and Hygiea are the largest among the families, and in combination with the analysis of the distribution of physical properties among their members, suggests that these families could have undergone a partial differentiation process. In contrast, Lixiaohua and Veritas present smaller parent body sizes, and are unlikely to have gone through such processing. The Ursula parent body is of intermediate size, and more data are required to understand the origin of the family. We estimated the density limits for the members of the Themis, Hygiea, and Ursula families, where the first two present densities compatible with C-type asteroids, and the latter with P-type, in agreement with their spectroscopic properties.
    
    The results of the analysis performed in this work suggest a diverse scenario for the formation of the parent bodies of the families studied. The differences between Themis, Hygiea, and Veritas could be related to the initial masses of their progenitors, while Ursula and Lixiaohua could have formed in separate and distinct regions of the protoplanetary disk, and were scattered to their current orbits. However, additional investigations of the compositions and physical properties of their members and further analyses of the remaining primitive outer-belt families are required to test these hypotheses. The continuation of the PRIMASS and future data releases of asteroid spectra from Gaia will greatly contribute to enhancing the analysis of these populations.

\begin{acknowledgements}
      M. De Pr\'a acknowledges funding from the Prominent Postdoctoral Program of the University of Central Florida. N. Pinilla-Alonso acknowledges support from the 17-PDART17-2-0097 project funded by the PDART-ROSES 2017 program. J. Licandro acknowledge support from the AYA2015-67772-R  (MINECO, Spain). J. de Le\'on acknowledges  support from MINECO under the 2015 Severo Ochoa Program SEV-2015-0548 and the AYA-2017-89090-P. This work is based in part on observations obtained at the Southern Astrophysical Research Telescope (SOAR), which is a joint project of the Minist\'erio da Ci\^encia, Tecnologia, Inova\c{c}\~ao e Comunica\c{c}\~oes (MCTIC) do Brasil, the U.S. National Optical Astronomy Observatory (NOAO), the University of North Carolina at Chapel Hill (UNC), and Michigan State University (MSU). 
\end{acknowledgements}


\bibliographystyle{aa}
\bibliography{export-bibtex}

\begin{thebibliography}{75}
\expandafter\ifx\csname natexlab\endcsname\relax\def\natexlab#1{#1}\fi

\bibitem[{{Astropy Collaboration} {et~al.}(2013){Astropy Collaboration},
  {Robitaille}, {Tollerud}, {Greenfield}, {Droettboom}, {Bray}, {Aldcroft},
  {Davis}, {Ginsburg}, {Price-Whelan}, {Kerzendorf}, {Conley}, {Crighton},
  {Barbary}, {Muna}, {Ferguson}, {Grollier}, {Parikh}, {Nair}, {Unther},
  {Deil}, {Woillez}, {Conseil}, {Kramer}, {Turner}, {Singer}, {Fox}, {Weaver},
  {Zabalza}, {Edwards}, {Azalee Bostroem}, {Burke}, {Casey}, {Crawford},
  {Dencheva}, {Ely}, {Jenness}, {Labrie}, {Lim}, {Pierfederici}, {Pontzen},
  {Ptak}, {Refsdal}, {Servillat}, \& {Streicher}}]{astropy:2013}
{Astropy Collaboration}, {Robitaille}, T.~P., {Tollerud}, E.~J., {et~al.} 2013,
  \aap, 558, A33

\bibitem[{{Bro{\v{z}}} {et~al.}(2013){Bro{\v{z}}}, {Morbidelli}, {Bottke},
  {Rozehnal}, {Vokrouhlick{\'y}}, \& {Nesvorn{\'y}}}]{2013A&A...551A.117B}
{Bro{\v{z}}}, M., {Morbidelli}, A., {Bottke}, W.~F., {et~al.} 2013, \aap, 551,
  A117

\bibitem[{{Bro{\v{z}}} \& {Vokrouhlick{\'y}}(2008)}]{2008MNRAS.390..715B}
{Bro{\v{z}}}, M. \& {Vokrouhlick{\'y}}, D. 2008, \mnras, 390, 715

\bibitem[{{Bus} \& {Binzel}(2004)}]{2004PDSS....1.....B}
{Bus}, S. \& {Binzel}, R.~P. 2004, NASA Planetary Data System, EAR

\bibitem[{{Bus} \& {Binzel}(2002)}]{bustaxonomy}
{Bus}, S.~J. \& {Binzel}, R.~P. 2002, \icarus, 158, 146

\bibitem[{{Campins} {et~al.}(2010){Campins}, {Hargrove}, {Pinilla-Alonso},
  {Howell}, {Kelley}, {Licandro}, {Moth{\'e}-Diniz}, {Fern{\'a}ndez}, \&
  {Ziffer}}]{2010Natur.464.1320C}
{Campins}, H., {Hargrove}, K., {Pinilla-Alonso}, N., {et~al.} 2010, \nat, 464,
  1320

\bibitem[{{Carbognani}(2017)}]{2017P&SS..147....1C}
{Carbognani}, A. 2017, \planss, 147, 1

\bibitem[{{Carruba}(2013)}]{2013MNRAS.431.3557C}
{Carruba}, V. 2013, \mnras, 431, 3557

\bibitem[{{Carry}(2012)}]{2012P&SS...73...98C}
{Carry}, B. 2012, \planss, 73, 98

\bibitem[{{Carvano} {et~al.}(2010){Carvano}, {Hasselmann}, {Lazzaro}, \&
  {Moth{\'e}-Diniz}}]{carvano2010}
{Carvano}, J.~M., {Hasselmann}, P.~H., {Lazzaro}, D., \& {Moth{\'e}-Diniz}, T.
  2010, \aap, 510, A43

\bibitem[{{Carvano} {et~al.}(2003){Carvano}, {Moth{\'e}-Diniz}, \&
  {Lazzaro}}]{2003Icar..161..356C}
{Carvano}, J.~M., {Moth{\'e}-Diniz}, T., \& {Lazzaro}, D. 2003, \icarus, 161,
  356

\bibitem[{{Castillo-Rogez} \& {Schmidt}(2010)}]{2010GeoRL..3710202C}
{Castillo-Rogez}, J.~C. \& {Schmidt}, B.~E. 2010, \grl, 37, L10202

\bibitem[{{Cellino} {et~al.}(2002){Cellino}, {Bus}, {Doressoundiram}, \&
  {Lazzaro}}]{2002aste.book..633C}
{Cellino}, A., {Bus}, S.~J., {Doressoundiram}, A., \& {Lazzaro}, D. 2002,
  {Spectroscopic Properties of Asteroid Families}, 633--643

\bibitem[{{de Le{\'o}n} {et~al.}(2012){de Le{\'o}n}, {Pinilla-Alonso},
  {Campins}, {Licand ro}, \& {Marzo}}]{2012Icar..218..196D}
{de Le{\'o}n}, J., {Pinilla-Alonso}, N., {Campins}, H., {Licand ro}, J., \&
  {Marzo}, G.~A. 2012, \icarus, 218, 196

\bibitem[{{de Le{\'o}n} {et~al.}(2016){de Le{\'o}n}, {Pinilla-Alonso}, {Delbo},
  {Campins}, {Cabrera-Lavers}, {Tanga}, {Cellino}, {Bendjoya}, {Gayon-Markt},
  {Licand ro}, {Lorenzi}, {Morate}, {Walsh}, {DeMeo}, {Landsman}, \&
  {Al{\'\i}-Lagoa}}]{2016Icar..266...57D}
{de Le{\'o}n}, J., {Pinilla-Alonso}, N., {Delbo}, M., {et~al.} 2016, \icarus,
  266, 57

\bibitem[{De~Pr{\'a} {et~al.}(2020)De~Pr{\'a}, Licandro, Pinilla-Alonso,
  Lorenzi, Rond{\'o}n, Carvano, Morate, \& De~Le{\'o}n}]{de2020spectroscopic}
De~Pr{\'a}, M., Licandro, J., Pinilla-Alonso, N., {et~al.} 2020, Icarus, 338,
  113473

\bibitem[{{De Pra} {et~al.}(2018){De Pra}, {Carvano}, {Morate},
  {Pinilla-Alonso}, \& {Licandro}}]{cana}
{De Pra}, M.~N., {Carvano}, J., {Morate}, D., {Pinilla-Alonso}, N., \&
  {Licandro}, J. 2018, in AAS/Division for Planetary Sciences Meeting
  Abstracts, 315.02

\bibitem[{{De Pr{\'a}} {et~al.}(2018){De Pr{\'a}}, {Pinilla-Alonso}, {Carvano},
  {Licandro}, {Campins}, {Moth{\'e}-Diniz}, {De Le{\'o}n}, \&
  {Al{\'\i}-Lagoa}}]{2018Icar..311...35D}
{De Pr{\'a}}, M.~N., {Pinilla-Alonso}, N., {Carvano}, J.~M., {et~al.} 2018,
  \icarus, 311, 35

\bibitem[{{DeMeo} {et~al.}(2015){DeMeo}, {Alexander}, {Walsh}, {Chapman}, \&
  {Binzel}}]{2015aste.book...13D}
{DeMeo}, F.~E., {Alexander}, C.~M.~O., {Walsh}, K.~J., {Chapman}, C.~R., \&
  {Binzel}, R.~P. 2015, {The Compositional Structure of the Asteroid Belt},
  13--41

\bibitem[{{DeMeo} {et~al.}(2009){DeMeo}, {Binzel}, {Slivan}, \&
  {Bus}}]{2009Icar..202..160D}
{DeMeo}, F.~E., {Binzel}, R.~P., {Slivan}, S.~M., \& {Bus}, S.~J. 2009,
  \icarus, 202, 160

\bibitem[{{DeMeo} \& {Carry}(2014)}]{2014Natur.505..629D}
{DeMeo}, F.~E. \& {Carry}, B. 2014, \nat, 505, 629

\bibitem[{{Descamps} {et~al.}(2007){Descamps}, {Marchis}, {Michalowski},
  {Vachier}, {Colas}, {Berthier}, {Assafin}, {Dunckel}, {Polinska}, {Pych},
  {Hestroffer}, {Miller}, {Vieira-Martins}, {Birlan}, {Teng-Chuen-Yu},
  {Peyrot}, {Payet}, {Dorseuil}, {L{\'e}onie}, \&
  {Dijoux}}]{2007Icar..187..482D}
{Descamps}, P., {Marchis}, F., {Michalowski}, T., {et~al.} 2007, \icarus, 187,
  482

\bibitem[{{Durda} {et~al.}(2007){Durda}, {Bottke}, {Nesvorn{\'y}}, {Enke},
  {Merline}, {Asphaug}, \& {Richardson}}]{2007Icar..186..498D}
{Durda}, D.~D., {Bottke}, W.~F., {Nesvorn{\'y}}, D., {et~al.} 2007, \icarus,
  186, 498

\bibitem[{{Feierberg} {et~al.}(1985){Feierberg}, {Lebofsky}, \&
  {Tholen}}]{1985Icar...63..183F}
{Feierberg}, M.~A., {Lebofsky}, L.~A., \& {Tholen}, D.~J. 1985, \icarus, 63,
  183

\bibitem[{{Florczak} {et~al.}(1999){Florczak}, {Lazzaro}, {Moth{\'e}-Diniz},
  {Angeli}, \& {Betzler}}]{1999A&AS..134..463F}
{Florczak}, M., {Lazzaro}, D., {Moth{\'e}-Diniz}, T., {Angeli}, C.~A., \&
  {Betzler}, A.~S. 1999, \aaps, 134, 463

\bibitem[{{Fornasier} {et~al.}(2014){Fornasier}, {Lantz}, {Barucci}, \&
  {Lazzarin}}]{2014Icar..233..163F}
{Fornasier}, S., {Lantz}, C., {Barucci}, M.~A., \& {Lazzarin}, M. 2014,
  \icarus, 233, 163

\bibitem[{{Fornasier} {et~al.}(2016){Fornasier}, {Lantz}, {Perna}, {Campins},
  {Barucci}, \& {Nesvorny}}]{2016Icar..269....1F}
{Fornasier}, S., {Lantz}, C., {Perna}, D., {et~al.} 2016, \icarus, 269, 1

\bibitem[{{Hasselmann} {et~al.}(2011){Hasselmann}, {Carvano}, \&
  {Lazzaro}}]{Hasselmann2011}
{Hasselmann}, P.~H., {Carvano}, J.~M., \& {Lazzaro}, D. 2011, NASA Planetary
  Data System, 145

\bibitem[{{Hendrix} \& {Vilas}(2019)}]{2019GeoRL..4614307H}
{Hendrix}, A.~R. \& {Vilas}, F. 2019, \grl, 46, 14,307

\bibitem[{{Hirayama}(1918)}]{1918AJ.....31..185H}
{Hirayama}, K. 1918, \aj, 31, 185

\bibitem[{{Hsieh} \& {Jewitt}(2006)}]{2006Sci...312..561H}
{Hsieh}, H.~H. \& {Jewitt}, D. 2006, Science, 312, 561

\bibitem[{{Hsieh} {et~al.}(2004){Hsieh}, {Jewitt}, \&
  {Fern{\'a}ndez}}]{2004AJ....127.2997H}
{Hsieh}, H.~H., {Jewitt}, D.~C., \& {Fern{\'a}ndez}, Y.~R. 2004, \aj, 127, 2997

\bibitem[{{Hsieh} {et~al.}(2018){Hsieh}, {Novakovi{\'c}}, {Kim}, \&
  {Brasser}}]{2018AJ....155...96H}
{Hsieh}, H.~H., {Novakovi{\'c}}, B., {Kim}, Y., \& {Brasser}, R. 2018, \aj,
  155, 96

\bibitem[{Hunter(2007)}]{matplotlib}
Hunter, J.~D. 2007, Computing in science \& engineering, 9, 90

\bibitem[{Jones {et~al.}(2001)Jones, Oliphant, Peterson, {et~al.}}]{scipy}
Jones, E., Oliphant, T., Peterson, P., {et~al.} 2001, {SciPy}: Open source
  scientific tools for {Python}, [Online; accessed <today>]

\bibitem[{{Kaluna} {et~al.}(2016){Kaluna}, {Masiero}, \&
  {Meech}}]{2016Icar..264...62K}
{Kaluna}, H.~M., {Masiero}, J.~R., \& {Meech}, K.~J. 2016, \icarus, 264, 62

\bibitem[{{Landsman} {et~al.}(2016){Landsman}, {Licandro}, {Campins}, {Ziffer},
  {Pr{\'a}}, \& {Cruikshank}}]{2016Icar..269...62L}
{Landsman}, Z.~A., {Licandro}, J., {Campins}, H., {et~al.} 2016, \icarus, 269,
  62

\bibitem[{{Lazzaro} {et~al.}(2007){Lazzaro}, {Angeli}, {Carvano},
  {Mothe-Diniz}, {Duffard}, \& {Florczak}}]{2007PDSS...51.....L}
{Lazzaro}, D., {Angeli}, C.~A., {Carvano}, J.~M., {et~al.} 2007, NASA Planetary
  Data System, EAR

\bibitem[{{Licandro} {et~al.}(2011){Licandro}, {Campins}, {Kelley}, {Hargrove},
  {Pinilla-Alonso}, {Cruikshank}, {Rivkin}, \& {Emery}}]{2011A&A...525A..34L}
{Licandro}, J., {Campins}, H., {Kelley}, M., {et~al.} 2011, \aap, 525, A34

\bibitem[{{Licandro} {et~al.}(2012){Licandro}, {Hargrove}, {Kelley}, {Campins},
  {Ziffer}, {Al{\'\i}-Lagoa}, {Fern{\'a}ndez}, \&
  {Rivkin}}]{2012A&A...537A..73L}
{Licandro}, J., {Hargrove}, K., {Kelley}, M., {et~al.} 2012, \aap, 537, A73

\bibitem[{{Mainzer} {et~al.}(2016){Mainzer}, {Bauer}, {Cutri}, {Grav},
  {Kramer}, {Masiero}, {Nugent}, {Sonnett}, {Stevenson}, \&
  {Wright}}]{mainzer2016}
{Mainzer}, A.~K., {Bauer}, J.~M., {Cutri}, R.~M., {et~al.} 2016, NASA Planetary
  Data System, 247

\bibitem[{{Marchis} {et~al.}(2008){Marchis}, {Descamps}, {Berthier},
  {Hestroffer}, {Vachier}, {Baek}, {Harris}, \&
  {Nesvorn{\'y}}}]{2008Icar..195..295M}
{Marchis}, F., {Descamps}, P., {Berthier}, J., {et~al.} 2008, \icarus, 195, 295

\bibitem[{{Marsset} {et~al.}(2016){Marsset}, {Vernazza}, {Birlan}, {DeMeo},
  {Binzel}, {Dumas}, {Milli}, \& {Popescu}}]{2016A&A...586A..15M}
{Marsset}, M., {Vernazza}, P., {Birlan}, M., {et~al.} 2016, \aap, 586, A15

\bibitem[{{Michel} {et~al.}(2011){Michel}, {Jutzi}, {Richardson}, \&
  {Benz}}]{2011Icar..211..535M}
{Michel}, P., {Jutzi}, M., {Richardson}, D.~C., \& {Benz}, W. 2011, \icarus,
  211, 535

\bibitem[{{Morate} {et~al.}(2018){Morate}, {de Le{\'o}n}, {De Pr{\'a}},
  {Licandro}, {Cabrera-Lavers}, {Campins}, \&
  {Pinilla-Alonso}}]{2018A&A...610A..25M}
{Morate}, D., {de Le{\'o}n}, J., {De Pr{\'a}}, M., {et~al.} 2018, \aap, 610,
  A25

\bibitem[{{Morate} {et~al.}(2016){Morate}, {de Le{\'o}n}, {De Pr{\'a}},
  {Licandro}, {Cabrera-Lavers}, {Campins}, {Pinilla-Alonso}, \&
  {Al{\'\i}-Lagoa}}]{2016A&A...586A.129M}
{Morate}, D., {de Le{\'o}n}, J., {De Pr{\'a}}, M., {et~al.} 2016, \aap, 586,
  A129

\bibitem[{{Morate} {et~al.}(2019){Morate}, {de Le{\'o}n}, {De Pr{\'a}},
  {Licandro}, {Pinilla-Alonso}, {Campins}, {Arredondo}, {Carvano}, {Lazzaro},
  \& {Cabrera-Lavers}}]{2019A&A...630A.141M}
{Morate}, D., {de Le{\'o}n}, J., {De Pr{\'a}}, M., {et~al.} 2019, \aap, 630,
  A141

\bibitem[{{Morbidelli} {et~al.}(2000){Morbidelli}, {Chambers}, {Lunine},
  {Petit}, {Robert}, {Valsecchi}, \& {Cyr}}]{2000M&PS...35.1309M}
{Morbidelli}, A., {Chambers}, J., {Lunine}, J.~I., {et~al.} 2000, Meteoritics
  and Planetary Science, 35, 1309

\bibitem[{{Moth{\'e}-Diniz} {et~al.}(2001){Moth{\'e}-Diniz}, {di Martino},
  {Bendjoya}, {Doressoundiram}, \& {Migliorini}}]{2001Icar..152..117M}
{Moth{\'e}-Diniz}, T., {di Martino}, M., {Bendjoya}, P., {Doressoundiram}, A.,
  \& {Migliorini}, F. 2001, \icarus, 152, 117

\bibitem[{{Moth{\'e}-Diniz} {et~al.}(2005){Moth{\'e}-Diniz}, {Roig}, \&
  {Carvano}}]{2005Icar..174...54M}
{Moth{\'e}-Diniz}, T., {Roig}, F., \& {Carvano}, J.~M. 2005, \icarus, 174, 54

\bibitem[{{Nesvorny}(2015)}]{2015PDSS..234.....N}
{Nesvorny}, D. 2015, NASA Planetary Data System, EAR

\bibitem[{{Nesvorn{\'y}} {et~al.}(2015){Nesvorn{\'y}}, {Bro{\v{z}}}, \&
  {Carruba}}]{2015aste.book..297N}
{Nesvorn{\'y}}, D., {Bro{\v{z}}}, M., \& {Carruba}, V. 2015, {Identification
  and Dynamical Properties of Asteroid Families}, 297--321

\bibitem[{{Popescu} {et~al.}(2016){Popescu}, {Licandro}, {Morate}, {de
  Le{\'o}n}, {Nedelcu}, {Rebolo}, {McMahon}, {Gonzalez-Solares}, \&
  {Irwin}}]{Popescu2016}
{Popescu}, M., {Licandro}, J., {Morate}, D., {et~al.} 2016, Astronomy and
  Astrophysics, 591, A115

\bibitem[{{Pravec} \& {Harris}(2000)}]{2000Icar..148...12P}
{Pravec}, P. \& {Harris}, A.~W. 2000, \icarus, 148, 12

\bibitem[{{Pravec} \& {Harris}(2007)}]{2007Icar..190..250P}
{Pravec}, P. \& {Harris}, A.~W. 2007, \icarus, 190, 250

\bibitem[{{Price-Whelan} {et~al.}(2018){Price-Whelan}, {Sip{\H{o}}cz},
  {G{\"u}nther}, {Lim}, {Crawford}, {Conseil}, {Shupe}, {Craig}, {Dencheva},
  {Ginsburg}, {VanderPlas}, {Bradley}, {P{\'e}rez-Su{\'a}rez}, {de Val-Borro},
  {Paper Contributors}, {Aldcroft}, {Cruz}, {Robitaille}, {Tollerud},
  {Coordination Committee}, {Ardelean}, {Babej}, {Bach}, {Bachetti}, {Bakanov},
  {Bamford}, {Barentsen}, {Barmby}, {Baumbach}, {Berry}, {Biscani}, {Boquien},
  {Bostroem}, {Bouma}, {Brammer}, {Bray}, {Breytenbach}, {Buddelmeijer},
  {Burke}, {Calderone}, {Cano Rodr{\'\i}guez}, {Cara}, {Cardoso}, {Cheedella},
  {Copin}, {Corrales}, {Crichton}, {D{\textquoteright}Avella}, {Deil},
  {Depagne}, {Dietrich}, {Donath}, {Droettboom}, {Earl}, {Erben}, {Fabbro},
  {Ferreira}, {Finethy}, {Fox}, {Garrison}, {Gibbons}, {Goldstein}, {Gommers},
  {Greco}, {Greenfield}, {Groener}, {Grollier}, {Hagen}, {Hirst}, {Homeier},
  {Horton}, {Hosseinzadeh}, {Hu}, {Hunkeler}, {Ivezi{\'c}}, {Jain}, {Jenness},
  {Kanarek}, {Kendrew}, {Kern}, {Kerzendorf}, {Khvalko}, {King}, {Kirkby},
  {Kulkarni}, {Kumar}, {Lee}, {Lenz}, {Littlefair}, {Ma}, {Macleod},
  {Mastropietro}, {McCully}, {Montagnac}, {Morris}, {Mueller}, {Mumford},
  {Muna}, {Murphy}, {Nelson}, {Nguyen}, {Ninan}, {N{\"o}the}, {Ogaz}, {Oh},
  {Parejko}, {Parley}, {Pascual}, {Patil}, {Patil}, {Plunkett}, {Prochaska},
  {Rastogi}, {Reddy Janga}, {Sabater}, {Sakurikar}, {Seifert}, {Sherbert},
  {Sherwood-Taylor}, {Shih}, {Sick}, {Silbiger}, {Singanamalla}, {Singer},
  {Sladen}, {Sooley}, {Sornarajah}, {Streicher}, {Teuben}, {Thomas},
  {Tremblay}, {Turner}, {Terr{\'o}n}, {van Kerkwijk}, {de la Vega}, {Watkins},
  {Weaver}, {Whitmore}, {Woillez}, {Zabalza}, \& {Contributors}}]{astropy:2018}
{Price-Whelan}, A.~M., {Sip{\H{o}}cz}, B.~M., {G{\"u}nther}, H.~M., {et~al.}
  2018, \aj, 156, 123

\bibitem[{{Rivkin}(2012)}]{2012Icar..221..744R}
{Rivkin}, A.~S. 2012, \icarus, 221, 744

\bibitem[{{Rivkin} {et~al.}(2014){Rivkin}, {Asphaug}, \&
  {Bottke}}]{2014Icar..243..429R}
{Rivkin}, A.~S., {Asphaug}, E., \& {Bottke}, W.~F. 2014, \icarus, 243, 429

\bibitem[{{Rivkin} {et~al.}(2015){Rivkin}, {Campins}, {Emery}, {Howell},
  {Licandro}, {Takir}, \& {Vilas}}]{2015aste.book...65R}
{Rivkin}, A.~S., {Campins}, H., {Emery}, J.~P., {et~al.} 2015, {Astronomical
  Observations of Volatiles on Asteroids}, 65--87

\bibitem[{{Rivkin} {et~al.}(2003){Rivkin}, {Davies}, {Johnson}, {Ellison},
  {Trilling}, {Brown}, \& {Lebofsky}}]{2003M&PS...38.1383R}
{Rivkin}, A.~S., {Davies}, J.~K., {Johnson}, J.~R., {et~al.} 2003, Meteoritics
  and Planetary Science, 38, 1383

\bibitem[{{Rivkin} \& {Emery}(2010)}]{2010Natur.464.1322R}
{Rivkin}, A.~S. \& {Emery}, J.~P. 2010, \nat, 464, 1322

\bibitem[{{Rivkin} {et~al.}(2019){Rivkin}, {Howell}, \&
  {Emery}}]{2019JGRE..124.1393R}
{Rivkin}, A.~S., {Howell}, E.~S., \& {Emery}, J.~P. 2019, Journal of
  Geophysical Research (Planets), 124, 1393

\bibitem[{{Takir} \& {Emery}(2012)}]{2012Icar..219..641T}
{Takir}, D. \& {Emery}, J.~P. 2012, \icarus, 219, 641

\bibitem[{{Tanga} {et~al.}(1999){Tanga}, {Cellino}, {Michel}, {Zappal{\`a}},
  {Paolicchi}, \& {Dell'Oro}}]{1999Icar..141...65T}
{Tanga}, P., {Cellino}, A., {Michel}, P., {et~al.} 1999, \icarus, 141, 65

\bibitem[{{Tedesco}(1994)}]{1994PDSS..101...96T}
{Tedesco}, E.~F. 1994, NASA Planetary Data System

\bibitem[{{Usui} {et~al.}(2019){Usui}, {Hasegawa}, {Ootsubo}, \&
  {Onaka}}]{2019PASJ...71....1U}
{Usui}, F., {Hasegawa}, S., {Ootsubo}, T., \& {Onaka}, T. 2019, \pasj, 71, 1

\bibitem[{Van Der~Walt {et~al.}(2011)Van Der~Walt, Colbert, \&
  Varoquaux}]{numpy}
Van Der~Walt, S., Colbert, S.~C., \& Varoquaux, G. 2011, Computing in Science
  \& Engineering, 13, 22

\bibitem[{{Vernazza} {et~al.}(2019){Vernazza}, {Jorda}, {{\v{S}}eve{\v{c}}ek},
  {Bro{\v{z}}}, {Viikinkoski}, {Hanu{\v{s}}}, {Carry}, {Drouard}, {Ferrais},
  {Marsset}, {Marchis}, {Birlan}, {Podlewska-Gaca}, {Jehin}, {Bartczak},
  {Dudzinski}, {Berthier}, {Castillo-Rogez}, {Cipriani}, {Colas}, {DeMeo},
  {Dumas}, {Durech}, {Fetick}, {Fusco}, {Grice}, {Kaasalainen}, {Kryszczynska},
  {Lamy}, {Le Coroller}, {Marciniak}, {Michalowski}, {Michel}, {Rambaux},
  {Santana-Ros}, {Tanga}, {Vachier}, {Vigan}, {Witasse}, {Yang}, {Gillon},
  {Benkhaldoun}, {Szakats}, {Hirsch}, {Duffard}, {Chapman}, \&
  {Maestre}}]{2019NatAs.tmp..477V}
{Vernazza}, P., {Jorda}, L., {{\v{S}}eve{\v{c}}ek}, P., {et~al.} 2019, Nature
  Astronomy, 477

\bibitem[{{Vilas}(1994)}]{1994Icar..111..456V}
{Vilas}, F. 1994, \icarus, 111, 456

\bibitem[{{Vilas}(1995)}]{1995Icar..115..217V}
{Vilas}, F. 1995, \icarus, 115, 217

\bibitem[{{Vilas} {et~al.}(2006){Vilas}, {Smith}, {McFadden}, {Gaffey},
  {Larson}, {Hatch}, \& {Jarvis}}]{2006PDSS...45.....V}
{Vilas}, F., {Smith}, B.~A., {McFadden}, L.~A., {et~al.} 2006, NASA Planetary
  Data System, EAR

\bibitem[{{Warner} {et~al.}(2009){Warner}, {Harris}, \&
  {Pravec}}]{2009Icar..202..134W}
{Warner}, B.~D., {Harris}, A.~W., \& {Pravec}, P. 2009, \icarus, 202, 134

\bibitem[{{Xu} {et~al.}(1996){Xu}, {Binzel}, {Burbine}, \&
  {Bus}}]{1996PDSS..168.....X}
{Xu}, S., {Binzel}, R.~P., {Burbine}, T.~H., \& {Bus}, S.~J. 1996, NASA
  Planetary Data System, EAR

\bibitem[{{Zappala} \& {Cellino}(1992)}]{1992CeMDA..54..207Z}
{Zappala}, V. \& {Cellino}, A. 1992, Celestial Mechanics and Dynamical
  Astronomy, 54, 207

\bibitem[{{Ziffer} {et~al.}(2011){Ziffer}, {Campins}, {Licandro}, {Walker},
  {Fernandez}, {Clark}, {Mothe-Diniz}, {Howell}, \&
  {Deshpande}}]{2011Icar..213..538Z}
{Ziffer}, J., {Campins}, H., {Licandro}, J., {et~al.} 2011, \icarus, 213, 538

\end{thebibliography}
%

%
%-------------------------------------------------------------
%          For the appendices, table longer than a single page
%-------------------------------------------------------------

% Table will be print automatically at the end of the document, 
% after the whole appendices
\onecolumn
\begin{appendix} %First appendix
\section{Additional figures and tables}
\label{appendix}
\setcounter{figure}{0}    
\setcounter{table}{0} 
\footnotesize

\footnotesize
\centering

        \begin{table*}[h!]
        \footnotesize
        \centering
        \caption{Asteroids observational conditions - visible}
        \setlength\tabcolsep{3.0pt}
        \label{table:visobs}
        \begin{tabular}{lllcccccccc}
                \hline
                Asteroid & Asteroid & Family & Date & Time Start & Slit & Airmass &  $V$ & $\alpha$ & T$_{EXP}$ & Solar\\
                Number & Name & Name &  & (UT) & ($"$) &   &  $(mag)$ & ($^\circ$) & (s) & Analogs\\
                
                \hline
                268 & Adorea & Themis & 2012-03-27 & 07:55 & 1.68 & 1.08 & 12.48 & 12.60 & 60.0 & 1,2,3,4 \\
                561 & Ingwelde & Themis &  2011-02-08 & 01:44 & 1.03 & 1.765 & 16.29 & 19.01 & 360.0 & 1 \\
                1581 & Abanderada & Themis &  2011-02-01 & 06:56 & 1.03 & 1.263 & 15.53 & 15.52 & 480.0 & 1 \\
                1674 & Groeneveld & Themis &  2011-02-08 & 02:07 & 1.03 & 1.97 & 16.00 & 18.51 & 360.0 & 1 \\
                1782 & Schneller & Themis &  2012-03-27 & 06:01 & 1.68 & 1.113 & 15.89 & 4.37 & 420.0 & 1,2,3,4 \\
                2534 & Houzeau & Themis & 2011-02-01 & 04:29 & 1.03 & 1.557 & 16.12 & 2.64 & 320.0 & 1 \\
                2563 & Boyarchuk & Themis &  2011-02-08 & 05:23 & 1.03 & 1.347 & 16.18 & 14.74 & 480.0 & 1 \\
                2918 & Salazar & Themis &  2012-03-29 & 04:32 & 1.68 & 1.14 & 16.55 & 1.29 & 480.0 & 1,3,5 \\
                5429 & 1988 BZ1 & Themis &  2011-01-31 & 06:14 & 1.03 & 1.497 & 16.55 & 2.48 & 540.0 & 1 \\
                \hline
                5594 & Jimmiller & Veritas & 2011-01-31 & 01:58 & 1.03 & 1.66 & 16.66 & 5.97 & 400.0 & 1 \\
                6343 & 1993 VK & Veritas & 2012-03-27 & 00:48 & 1.68 & 1.48 & 17.83 & 12.06 & 600.0 & 1,2,3,4 \\
                6374 & Beslan & Veritas & 2011-02-01 & 02:44 & 1.03 & 1.36 & 17.06 & 8.33 & 600.0 & 1 \\
                7231 & Porco & Veritas & 2012-03-29 & 07:07 & 1.68 & 1.2 & 17.86 & 17.97 & 600.0 & 1,3,5 \\
                9860 & Archaeopteryx &Veritas & 2012-03-29 & 06:02 & 1.68 & 1.29 & 18.03 & 2.63 & 600.0 & 1,3,5 \\
                13537 & 1991 SG & Veritas &2012-03-29 & 08:21 & 1.68 & 1.253 & 18.07 & 8.90 & 720.0 & 1,3,5 \\
                14264 & 2000 AH142 &Veritas & 2012-03-28 & 04:01 & 1.68 & 1.18 & 17.90 & 1.81 & 600.0 & 1,4,5 \\
                24638 & 1981 UC23 &Veritas & 2012-03-28 & 05:19 & 1.68 & 1.11 & 17.71 & 1.46 & 600.0 & 1,4,5 \\
                45378 & 2000 AD118 &Veritas & 2012-03-27 & 02:12 & 1.68 & 1.15 & 17.32 & 4.81 & 600.0 & 1,2,3,4 \\
                \hline
                1209 & Pumma & Hygiea & 2012-03-27 & 08:13 & 1.68 & 1.14 & 15.29 & 11.73 & 360.0 & 1,2,3,4 \\
                3626 & Ohsaki & Hygiea & 2012-03-27 & 03:48 & 1.68 & 1.313 & 17.39 & 10.46 & 600.0 & 1,2,3,4 \\
                5116 & Korsor & Hygiea & 2011-02-08 & 07:16 & 1.03 & 1.16 & 17.47 & 14.76 & 540.0 & 1 \\
                5794 & Irmina & Hygiea &2011-02-01 & 01:47 & 1.03 & 1.473 & 17.57 & 12.08 & 600.0 & 1 \\
                6415 & 1993 VR3 & Hygiea &2012-03-27 & 04:46 & 1.68 & 1.123 & 17.05 & 2.01 & 600.0 & 1,2,3,4 \\
                15134 & 2000 ED92 & Hygiea &2011-02-08 & 03:01 & 1.03 & 1.402 & 17.34 & 3.35 & 495.0 & 1 \\
                23916 & 1998 SD131 & Hygiea &2012-03-27 & 03:03 & 1.68 & 1.203 & 17.43 & 0.52 & 600.0 & 1,2,3,4 \\
                24358 & 2000 AV117 & Hygiea &2012-03-28 & 02:35 & 1.68 & 1.323 & 17.30 & 4.53 & 600.0 & 1,4,5 \\
                24956 & 1997 SN10 & Hygiea &2012-03-29 & 03:16 & 1.68 & 1.192 & 17.86 & 2.16 & 600.0 & 1,3,5 \\
                25829 & 2000 DU108 & Hygiea &2012-03-27 & 08:44 & 1.68 & 1.29 & 17.72 & 9.08 & 600.0 & 1,2,3,4 \\
                26719 & 2001 HQ5 & Hygiea &2012-03-29 & 02:16 & 1.68 & 1.29 &  17.68 & 2.62 & 600.0 & 1,3,5 \\
                \hline
                \multicolumn{9}{l}{\footnotesize Solar Analogs: (1) L102-1081, (2) L107-684, (3) L107-998, (4) HD44594, (5) HD144584}
        \end{tabular}   
\end{table*}

        \begin{figure*}[h!]
        \centering
        \includegraphics[scale=0.6]{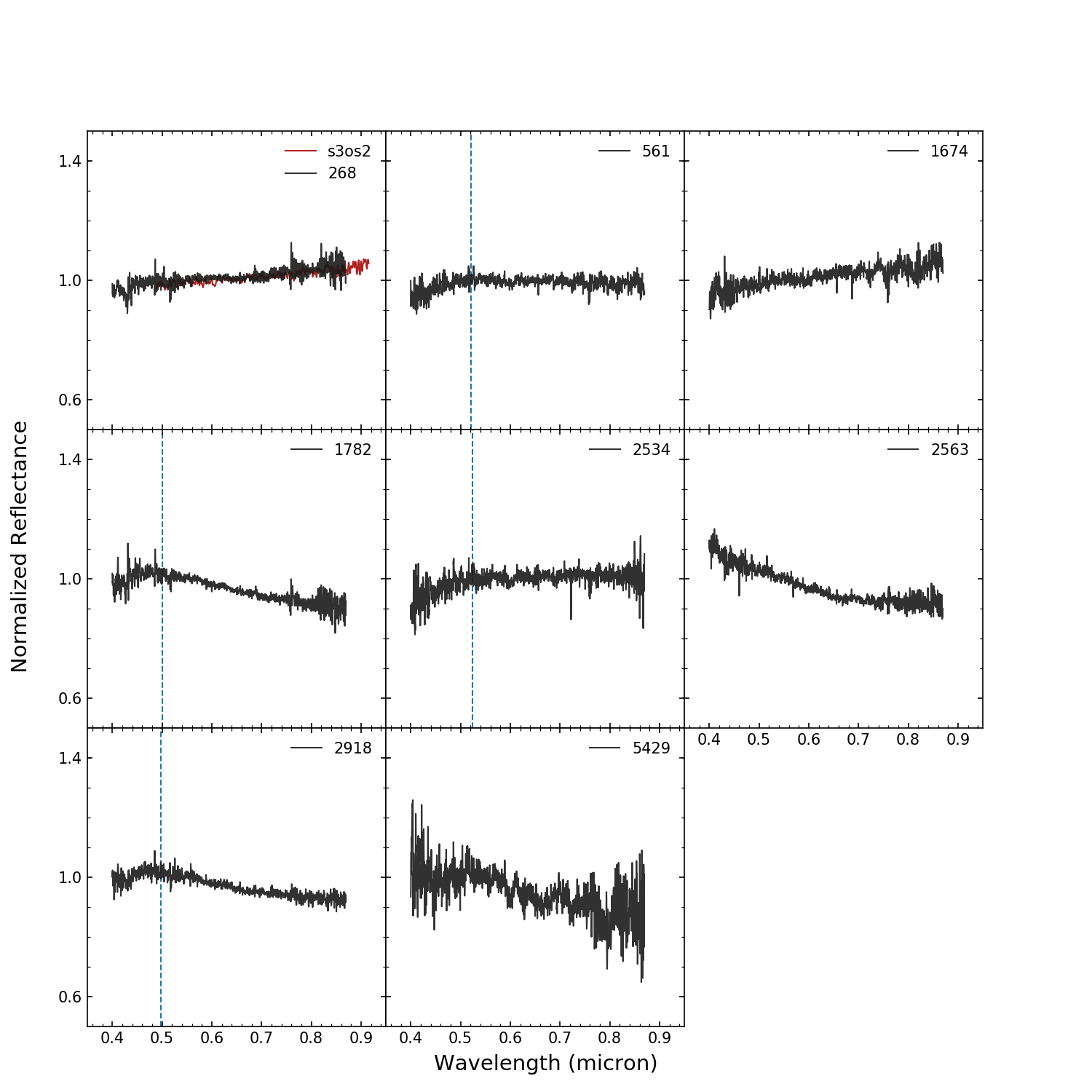}
        \caption{Visible spectroscopy of asteroids from the Themis family. Spectra of the same objects that were observed by different works are shown in red. The blue lines represent the turn-off point.}
        \label{fig:themis}
        \end{figure*}
        
        \begin{figure*}[h!]
        \centering
        \includegraphics[scale=0.6]{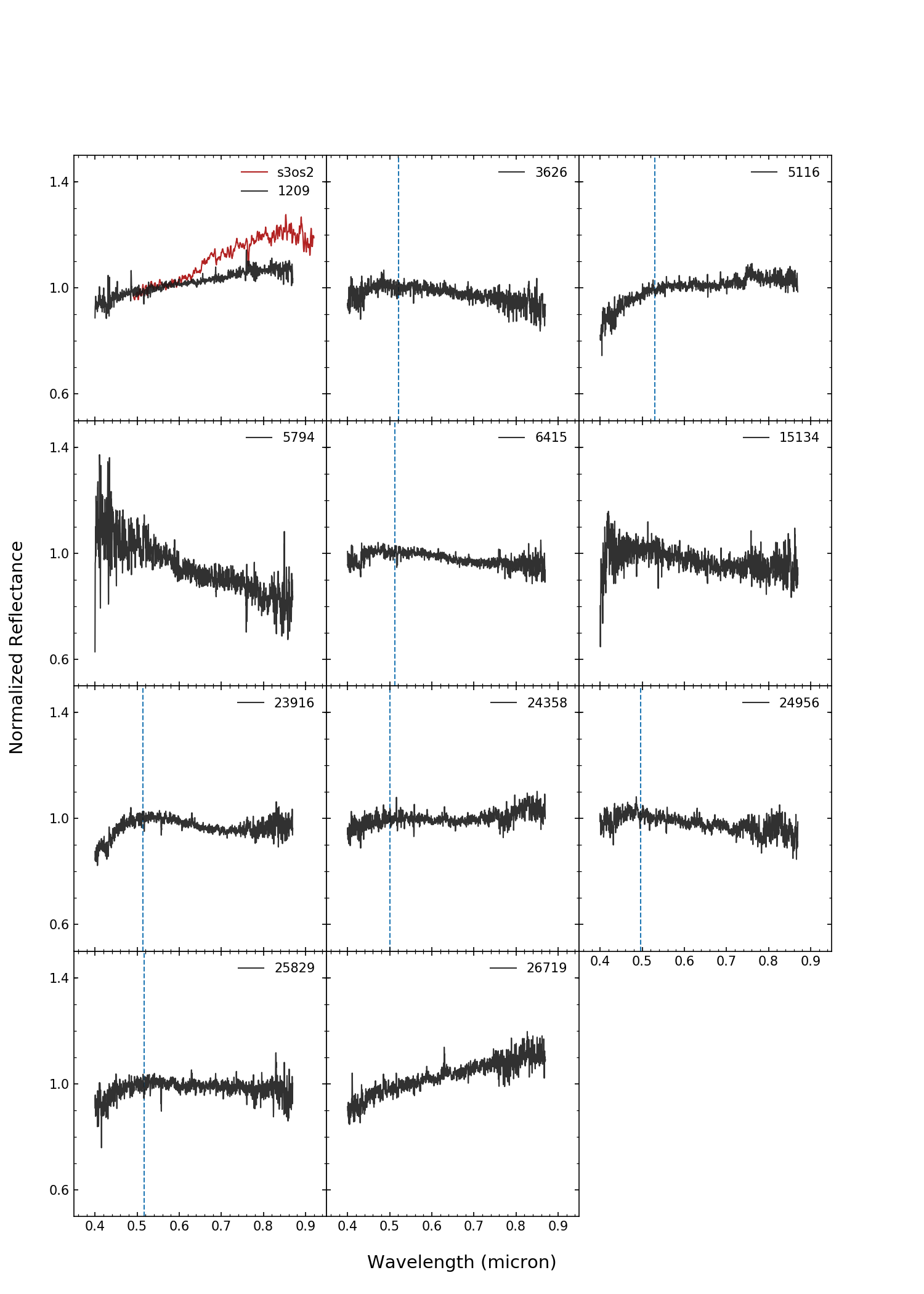}
        \caption{Visible spectroscopy of asteroids from the Hygiea family. Spectra of the same objects that were observed by different works are shown in red. The blue lines represent the turn-off point.}
        \label{fig:hygiea}
        \end{figure*}

        \begin{figure*}[h!]
        \centering
        \includegraphics[scale=0.6]{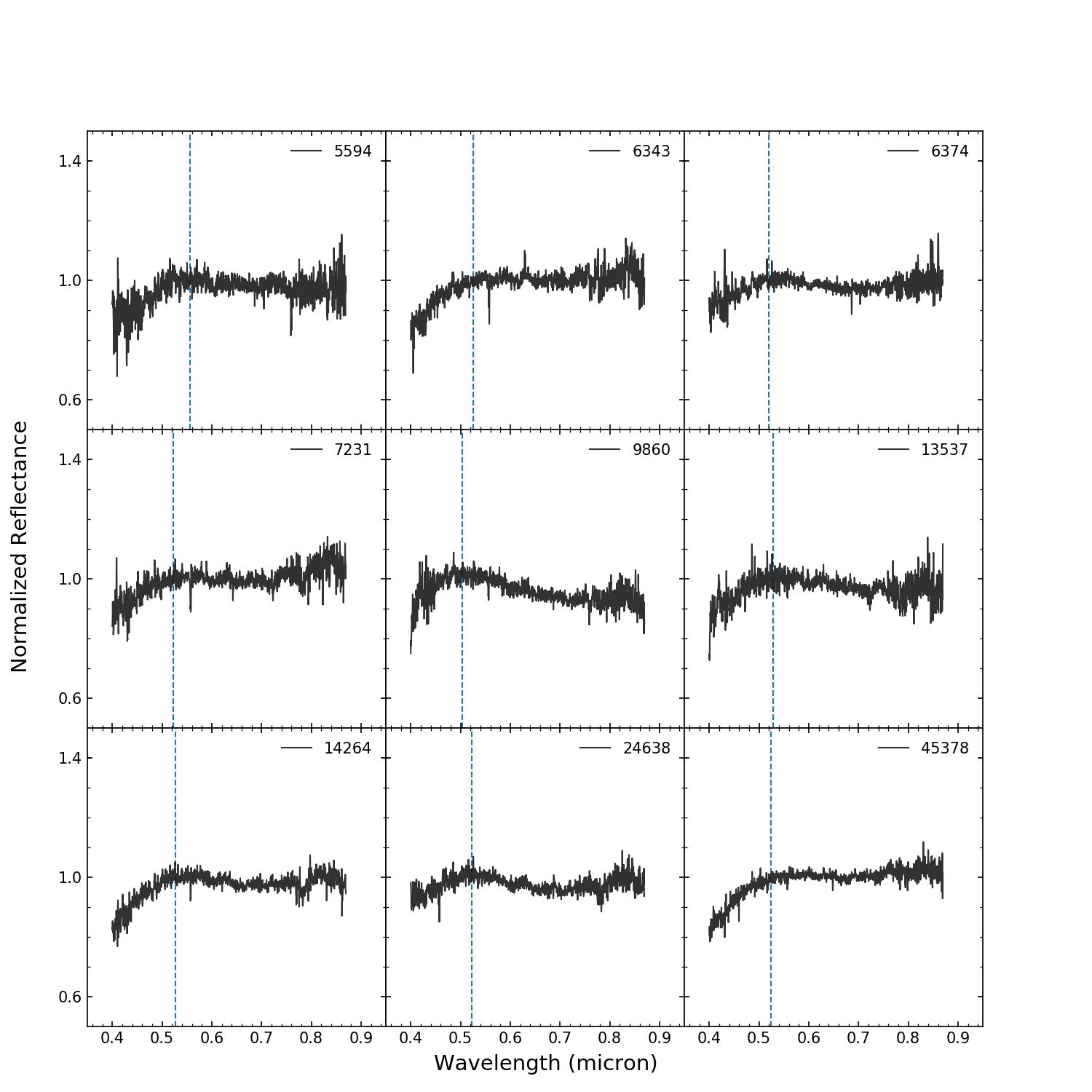}
        \caption{Visible spectroscopy of asteroids from the Veritas family. The blue lines represent the turn-off point.}
        \label{fig:veritas}
        \end{figure*}

\end{appendix}

%-------------------------------------------------------------
%                   For appendices and landscape, large table:
%                    in the preamble, use: \usepackage{lscape}
%-------------------------------------------------------------

\clearpage
\onecolumn

% \begin{table}
% \centering
{\setlength{\tabcolsep}{2pt}
        \begin{longtable}{llcccccccccl}
        \caption{Results for the spectroscopic parametrization. ($^1$) Reference labels are: 1- This work, 2- SMASS \citep{1996PDSS..168.....X}, 3-SMASSII \citep{2004PDSS....1.....B}, 4-S3OS2 \citep{2007PDSS...51.....L}, 5 - \cite{2006PDSS...45.....V}, 6 - \cite{2001Icar..152..117M}. The symbols  '-' indicate the spectra that do not show the feature and '*' where the wavelength coverage was not appropriate to evaluate the presence of the feature.}
        \label{tab:results}\\
        \hline
        Asteroid &  Family &  Taxonomic &  Visible &  Visible &  Turn-off & Turn-off &  0.7 $\mu m$ &   0.7 $\mu m$ &  0.7 $\mu m$ &  0.7 $\mu m$ &      ref$^1$ \\
        Number & & class & slope & slope error & point & point error & band depth & band depth & band center & band center & \\
        & & & ($\%/1000$\AA) &  ($\%/1000$\AA) & ($\AA$) & ($\AA$) & (\%) & error (\%) & ($\AA$) & error ($\AA$)\\
        \hline
        10    &  Hygiea &   B &  -2.191 &      1.000 &  6087 &       18 &      - &         - &      - &          - &    2 \\
    &  Hygiea &   C &   0.133 &      1.000 &  6010 &       39 &      - &         - &      - &          - &    3 \\
    &  Hygiea &   C &  -0.072 &      0.031 &     - &        - &      - &         - &      - &          - &    6 \\
52    &  Hygiea &  Cg &   2.227 &      1.001 &  6039 &       50 &      - &         - &      - &          - &    3 \\
100   &  Hygiea &   S &   7.272 &      1.001 &     * &        * &      - &         - &      - &          - &    3 \\
   &  Hygiea &   A &  11.771 &      0.165 &     - &        - &      - &         - &      - &          - &    6 \\
108   &  Hygiea &  Sv &   8.011 &      1.002 &     * &        * &      - &         - &      - &          - &    3 \\
159   &  Hygiea &  Ch &  -0.098 &      1.000 &  5345 &       15 &  2.294 &     0.102 &   6964 &         44 &    3 \\
538   &  Hygiea &   B &  -0.558 &      0.046 &     - &        - &      - &         - &      - &          - &    6 \\
867   &  Hygiea &   B &  -3.486 &      0.114 &    -  &        - &      - &         - &      - &          - &    6 \\
1107  &  Hygiea &  Xk &   2.782 &      1.000 &     - &        - &      - &         - &      - &          - &    3 \\
 1109  &  Hygiea &   D &   7.143 &      1.000 &     * &        * &      - &         - &      - &          - &    4 \\
  &  Hygiea &   D &   7.197 &      0.106 &     - &        - &      - &         - &      - &          - &    6 \\
1209  &  Hygiea &   X &   2.813 &      0.020 &     - &        - &      - &         - &      - &          - &    1 \\
  &  Hygiea &   T &   6.663 &      1.001 &     * &        * &      - &         - &      - &          - &    4 \\
  &  Hygiea &   T &   6.367 &      0.208 &     - &        - &      - &         - &      - &          - &    6 \\
1599  &  Hygiea &   S &   0.618 &      0.199 &     - &        - &      - &         - &      - &          - &    6 \\
2436  &  Hygiea &   B &  -3.540 &      0.231 &     - &        - &      - &         - &      - &          - &    6 \\
3626  &  Hygiea &   B &  -1.470 &      0.027 &  5212 &       30 &      - &         - &      - &          - &    1 \\
4667  &  Hygiea &   B &  -4.554 &      0.108 &     - &        - &      - &         - &      - &          - &    6 \\
4955  &  Hygiea &   B &  -3.420 &      1.000 &     * &        * &      - &         - &      - &          - &    4 \\
5116  &  Hygiea &   C &   2.934 &      0.022 &  5297 &       12 &      - &         - &      - &          - &    1 \\
5155  &  Hygiea &   C &  -0.666 &      0.079 &     - &        - &      - &         - &      - &          - &    6 \\
5794  &  Hygiea &   B &  -6.217 &      0.065 &     - &        - &      - &         - &      - &          - &    1 \\
6415  &  Hygiea &   B &  -1.050 &      0.021 &  5125 &       21 &      - &         - &      - &          - &    1 \\
15134 &  Hygiea &   B &  -1.686 &      0.047 &     - &        - &      - &         - &      - &          - &    1 \\
23916 &  Hygiea &   C &   0.297 &      0.023 &  5143 &        7 &  3.148 &     0.095 &   7119 &         21 &    1 \\
24358 &  Hygiea &  Cb &   1.118 &      0.024 &  5013 &       22 &  2.361 &     0.095 &   7058 &         34 &    1 \\
24956 &  Hygiea &   B &  -1.476 &      0.027 &  4967 &       20 &      - &         - &      - &          - &    1 \\
25829 &  Hygiea &   C &   0.304 &      0.032 &  5175 &       21 &      - &         - &      - &          - &    1 \\
26719 &  Hygiea &  Xc &   4.151 &      0.031 &     - &        - &      - &         - &      - &          - &    1 \\
        \hline
        24   &  Themis &   B & -0.852 &      0.021 &     - &        - &      - &         - &      - &          - &  3 \\
             &  Themis &   B & -0.808 &      0.022 &     * &        * &      - &         - &      - &          - &    4 \\
        62   &  Themis &   B & -1.790 &      0.037 &     5432 &        37 &  1.062 &     0.088 &   6741 &         50 &  3 \\
             &  Themis &   B & -2.074 &      0.014 &     * &        * &      - &         - &   6741 &         50 &    4 \\
        90   &  Themis &  Ch & -0.452 &      0.015 &  * &       * &   1.31 &     0.055 &   7087 &         28 &    4 \\
             &  Themis &   C &  0.409 &      0.024 &  6735 &       70 &      - &         - &   7087 &         28 &  3 \\
        171  &  Themis &   C &  0.269 &      0.017 &  * &       * &      - &         - &      - &          - &    4 \\
             &  Themis &   C &  0.666 &      0.022 &  5986 &       38 &      - &         - &      - &          - &  3 \\
        223  &  Themis &  Cb &  2.015 &      0.017 &     * &        * &      - &         - &      - &          - &    4 \\
        268  &  Themis &  Cb &  1.519 &      0.017 &     * &        * &      - &         - &      - &          - &    4 \\
             &  Themis &   C &  1.567 &      0.022 &     - &        - &      - &         - &      - &          - &        1 \\
        316  &  Themis &   C & -0.195 &      0.025 &     * &        * &      - &         - &      - &          - &    4 \\
        379  &  Themis &   C &  0.142 &      0.020 &  5728 &       39 &      - &         - &      - &          - &  3 \\
        383  &  Themis &   B & -1.918 &      0.055 &  5319 &       36 &      - &         - &      - &          - &  3 \\
        461  &  Themis &   X &  3.837 &      0.025 &     * &        * &      - &         - &      - &          - &    4 \\
        468  &  Themis &   C &  2.785 &      0.050 &     * &        * &      - &         - &      - &          - &    4 \\
        515  &  Themis &  Cb &  0.758 &      0.065 &     - &        - &      - &         - &      - &          - &  3 \\
        526  &  Themis &  Cb &  0.336 &      0.020 &     * &        * &      - &         - &      - &          - &    4 \\
        561  &  Themis &  Ch &  0.304 &      0.020 &  5219 &       29 &      - &         - &      - &          - &        1 \\
        621  &  Themis &   B & -0.905 &      0.034 &     * &        * &      - &         - &      - &          - &    4 \\
        767  &  Themis &   B & -2.318 &      0.031 &     - &        - &      - &         - &      - &          - &  3 \\
        846  &  Themis &  Cb &  0.912 &      0.026 &     * &        * &      - &         - &      - &          - &    4 \\
        848  &  Themis &  Cb &  0.619 &      0.019 &     * &        * &      - &         - &      - &          - &    4 \\
        936  &  Themis &   B & -1.363 &      0.014 &     * &        * &      - &         - &      - &          - &    4 \\
        954  &  Themis &  Cb &  0.783 &      0.028 &     * &        * &      - &         - &      - &          - &    4 \\
        981  &  Themis &   C & -0.002 &      0.018 &     * &        * &      - &         - &      - &          - &    4 \\
        1003 &  Themis &   B & -1.980 &      0.040 &     * &        * &      - &         - &      - &          - &    4 \\
        1229 &  Themis &   B & -1.779 &      0.040 &     * &        * &      - &         - &      - &          - &    4 \\
        1302 &  Themis &   B & -4.068 &      0.020 &     * &        * &  2.497 &     0.067 &   6596 &         54 &    2 \\
        1340 &  Themis &  Ch & -0.974 &      0.018 &     * &        * &  1.448 &     0.055 &   7084 &        126 &    4 \\
        1487 &  Themis &   B & -1.431 &      0.021 &     * &        * &      - &         - &      - &          - &    4 \\
        1539 &  Themis &   B & -1.726 &      0.027 &  5356 &        32 &      - &         - &      - &          - &  3 \\
             &  Themis &   B & -0.612 &      0.016 &     * &        * &      - &         - &      - &          - &    4 \\
        1576 &  Themis &   B & -1.713 &      0.014 &     * &        * &      - &         - &      - &          - &    4 \\
        1615 &  Themis &   B & -1.198 &      0.014 &     * &        * &      - &         - &      - &          - &    4 \\
        1674 &  Themis &   C &  2.061 &      0.027 &     - &        - &      - &         - &      - &          - &        1 \\
        1691 &  Themis &  Cb &  0.396 &      0.022 &     * &        * &      - &         - &      - &          - &    4 \\
        1782 &  Themis &   B & -2.655 &      0.023 &  5013 &       16 &      - &         - &      - &          - &        1 \\
        2296 &  Themis &  Cb &  0.328 &      0.023 &     * &        * &      - &         - &      - &          - &    4 \\
        2489 &  Themis &  Cb &  0.608 &      0.017 &     * &        * &  1.343 &     0.053 &   6531 &         53 &    4 \\
        2519 &  Themis &   B & -1.495 &      0.028 &     * &        * &      - &         - &      - &          - &    4 \\
        2524 &  Themis &  Ch & -0.765 &      0.015 &     * &        * &      - &         - &      - &          - &    4 \\
        2525 &  Themis &  Ch & -0.760 &      0.017 &     * &        * &  1.432 &     0.059 &   6848 &         43 &    4 \\
        2534 &  Themis &   C &  1.340 &      0.033 &  5241 &       42 &      - &         - &      - &          - &        1 \\
        2563 &  Themis &   B & -3.897 &      0.020 &     - &        - &  2.994 &     0.075 &   6804 &         28 &        1 \\
        2659 &  Themis &   B & -1.276 &      0.026 &  5633 &       35 &      - &         - &      - &          - &  3 \\
        2708 &  Themis &   B & -2.543 &      0.088 &     - &        - &      - &         - &      - &          - &  3 \\
        2918 &  Themis &   B & -2.225 &      0.016 &  4977 &       12 &  1.444 &     0.055 &   6963 &         44 &        1 \\
        3128 &  Themis &  Cb &  0.280 &      0.023 &     * &        * &      - &         - &      - &          - &    4 \\
        3179 &  Themis &  Ch &  0.811 &      0.215 &     - &        - &      - &         - &      - &          - &  3 \\
        3274 &  Themis &   C &  0.478 &      0.032 &     * &        * &      - &         - &      - &          - &    4 \\
        3507 &  Themis &   B & -0.978 &      0.019 &     * &        * &      - &         - &      - &          - &    4 \\
             &  Themis &  Ch &  0.423 &      0.127 &     - &        - &      - &         - &      - &          - &  3 \\
        3615 &  Themis &  Cb &  0.810 &      0.024 &     * &        * &      - &         - &      - &          - &    4 \\
        3663 &  Themis &   C &  0.067 &      0.026 &     * &        * &      - &         - &      - &          - &    4 \\
        3832 &  Themis &  Cb &  0.547 &      0.028 &     * &        * &      - &         - &      - &          - &    4 \\
        4143 &  Themis &  Cb &  0.326 &      0.023 &     * &        * &      - &         - &      - &          - &    4 \\
        4759 &  Themis &  Ch & -0.293 &      0.027 &     * &        * &      - &         - &      - &          - &    4 \\
        4778 &  Themis &   X &  3.630 &      0.047 &     * &        * &      - &         - &      - &          - &    4 \\
        5045 &  Themis &  Cb &  0.100 &      0.034 &     * &        * &      - &         - &      - &          - &    4 \\
        5429 &  Themis &   B & -3.854 &      0.063 &     - &        - &      - &         - &      - &          - &        1 \\
        6297 &  Themis &   B & -0.147 &      0.031 &     * &        * &      - &         - &      - &          - &    4 \\
        8518 &  Themis &  Ch & -0.742 &      0.024 &     * &        * &      - &         - &      - &          - &    4 \\
        8906 &  Themis &  Ch & -0.928 &      0.059 &     * &        * &      - &         - &      - &          - &    4 \\
        \hline
        490   &  Veritas &  Cgh &  1.745 &      0.024 &  5434 &       18 &  1.036 &     0.073 &   6902 &         38 &  3 \\
        1086  &  Veritas &   Ch &  0.081 &      0.022 &  * &       * &  4.424 &     0.073 &   7059 &         42 &    4 \\
              &  Veritas &   Ch & -0.248 &      0.051 &  5262 &       21 &  2.195 &      0.13 &   7059 &         42 &  3 \\
        2147  &  Veritas &   Ch &  1.213 &      0.106 &  5421 &       57 &   3.04 &     0.233 &   7276 &         57 &  3 \\
        2428  &  Veritas &   Ch &  0.039 &      0.099 &  5459 &       32 &    2.4 &     0.309 &   6959 &         77 &  3 \\
        2934  &  Veritas &    B & -1.175 &      0.128 &     - &        - &  3.634 &     0.448 &   6984 &        119 &  3 \\
        3090  &  Veritas &   Cg &  2.568 &      0.154 &  5738 &       50 &      - &         - &      - &          - &  3 \\
        3542  &  Veritas &    B & -0.462 &      0.152 &     - &        - &      - &         - &      - &          - &  3 \\
        5592  &  Veritas &    X &  2.792 &      0.040 &     * &        * &   3.32 &     0.095 &   7046 &         16 &    4 \\
        5594  &  Veritas &   Ch &  1.384 &      0.048 &  5558 &       20 &      - &         - &      - &          - &        1 \\
        6343  &  Veritas &    C &  2.355 &      0.035 &  5258 &       13 &      - &         - &      - &          - &        1 \\
        6374  &  Veritas &   Ch &  0.789 &      0.033 &  5201 &       18 &   2.99 &     0.117 &   7023 &         44 &        1 \\
        7231  &  Veritas &    C &  2.256 &      0.037 &  5222 &       17 &  3.351 &     0.134 &   7086 &         32 &        1 \\
        9860  &  Veritas &    B & -1.464 &      0.031 &  5030 &       12 &   3.32 &     0.122 &   7028 &         28 &        1 \\
        13537 &  Veritas &   Ch &  0.299 &      0.040 &  5284 &       19 &   1.74 &    0.1450 &   7159 &         49 &        1 \\
        14264 &  Veritas &   Ch &  1.582 &      0.026 &  5277 &        7 &  2.784 &     0.095 &   7056 &         23 &        1 \\
        24638 &  Veritas &   Ch &  0.144 &      0.025 &  5228 &       11 &  4.041 &       0.1 &   6992 &         21 &        1 \\
        45378 &  Veritas &  Cgh &  2.723 &      0.022 &  5241 &        9 &  1.567 &     0.078 &   7032 &         25 &        1 \\
        \hline
        375  &  Ursula &   X &  3.432 &      0.022 &  * &       * &      - &         - &   - &         - &    5 \\
             &  Ursula &  Xk &  3.175 &      0.048 &  6164 &       48 &      - &         - &   - &         - &  3 \\
        601  &  Ursula &   C &  1.304 &      0.076 &     6267 &        16 &      - &         - &      - &          - &  3 \\
             &  Ursula &   C &  0.215 &      0.015 &     * &        * &      - &         - &      - &          - &    4 \\
        618  &  Ursula &  Cb &  1.558 &      0.012 &     * &        * &      - &         - &      - &          - &    4 \\
        973  &  Ursula &   X &  3.380 &      0.032 &  * &       * &      - &         - &      - &          - &    4 \\
             &  Ursula &  Xk &  4.208 &      0.067 &  5892 &       80 &      - &         - &      - &          - &  3 \\
        1306 &  Ursula &   T &  5.510 &      0.030 &     * &        * &      - &         - &      - &          - &    4 \\
        1520 &  Ursula &  Cg &  1.979 &      0.069 &  5848 &      130 &      - &         - &      - &          - &  3 \\
        5959 &  Ursula &   X &  2.368 &      0.035 &     * &        * &      - &         - &      - &          - &    4 \\
        
        \hline
        \end{longtable}}
% \end{table}

\end{document}